\documentclass[acmtog]{acmart}

\setcopyright{none}

\usepackage{booktabs,multirow}  
\usepackage{amsfonts}           
\usepackage{amsmath}            
\usepackage{siunitx}
\usepackage{colortbl}       
\usepackage{xcolor}    
\usepackage[capitalise]{cleveref}

\acmJournal{TOG}

\begin{document}

\title{Real-time Global Illumination for Dynamic 3D Gaussian Scenes}




\author{Chenxiao Hu}
\affiliation{%
  \institution{Peking University}
  \city{Beijing}
  \country{China}}

\author{Meng Gai}
\affiliation{%
  \institution{Peking University}
  \city{Beijing}
  \country{China}}
\email{gaimeng@pku.edu.cn}

\author{Guoping Wang}
\affiliation{%
  \institution{Peking University}
  \city{Beijing}
  \country{China}}
\email{wgp@pku.edu.cn}
\orcid{0000-0001-7819-0076}

\author{Sheng Li}
\authornote{Sheng Li is the corresponding author. }
\affiliation{%
  \institution{Peking University}
  \city{Beijing}
  \country{China}}
\email{lisheng@pku.edu.cn}
\orcid{0000-0002-8901-2184}

\begin{abstract}

We present a real-time global illumination approach along with a pipeline for dynamic 3D Gaussian models and meshes.
Building on a formulated surface light transport model for 3D Gaussians, we address key performance challenges with a fast compound stochastic ray-tracing algorithm and an optimized 3D Gaussian rasterizer. Our pipeline integrates multiple real-time techniques to accelerate performance and achieve high-quality lighting effects. 
Our approach enables real-time rendering of dynamic scenes with interactively editable materials and dynamic lighting of diverse multi-lights settings, capturing mutual multi-bounce light transport (indirect illumination) between 3D Gaussians and mesh.
Additionally, we present a real-time renderer with an interactive user interface, validating our approach and demonstrating its practicality and high efficiency with over 40 fps in scenes including both 3D Gaussians and mesh. Furthermore, our work highlights the potential of 3D Gaussians in real-time applications with dynamic lighting, offering insights into performance and optimization.
\end{abstract}


\begin{teaserfigure}
    \centering
    \includegraphics[width=0.9\textwidth, trim={0.5cm 0.5cm 0.5cm 0.5cm}]{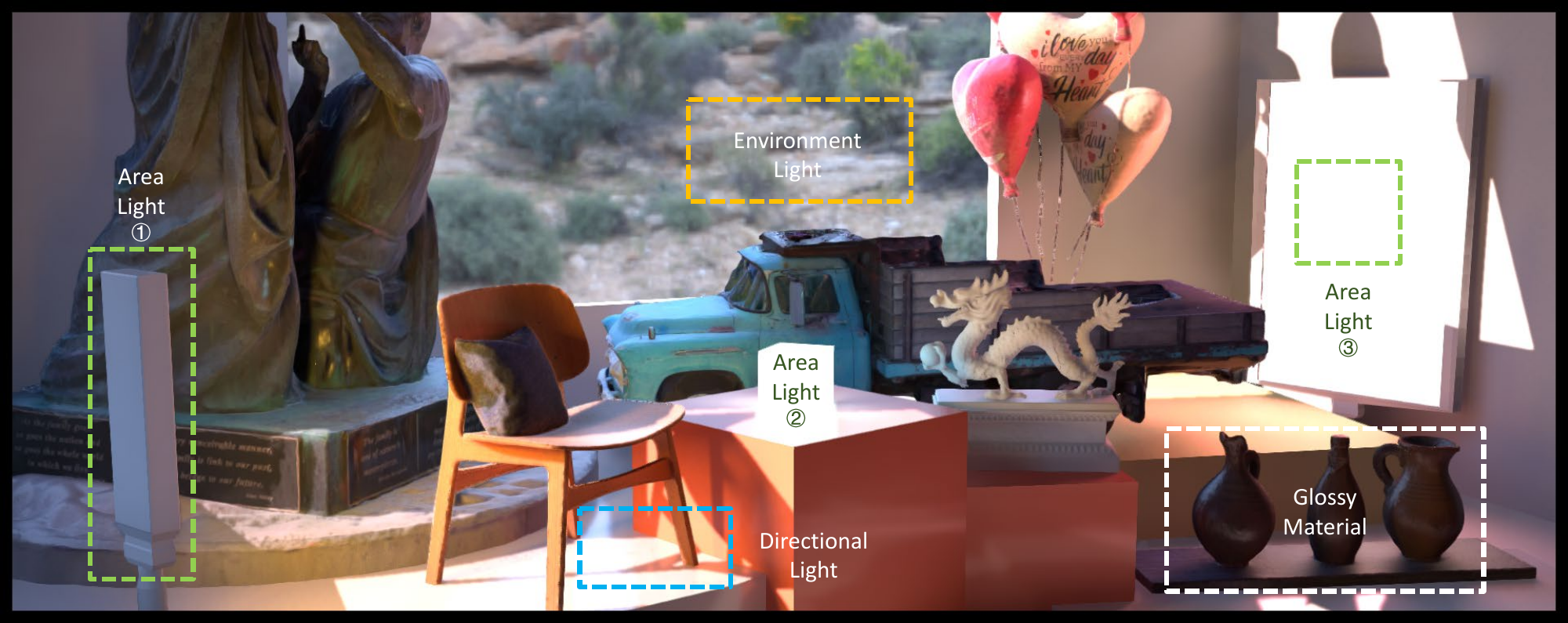}
    \caption{Global illumination by our renderer running at 1920x1088 resolution with over 40 fps. The compound scene includes five 3D Gaussian models (over 1.5 million on-screen 3D Gaussian primitives) and other mesh models (e.g., Chinese dragon). Our global illumination method can efficiently handle dynamic 3D Gaussians scenes, glossy materials, and multiple diverse dynamic light sources in real time, eliminating the need for pre-computation.}
    \label{fig:teaser}
\end{teaserfigure}

\maketitle

\section{Introduction}
3D Gaussians splitting (3DGS) \cite{kerbl20233dgs} has emerged as a promising technique to model 3D objects for efficient 3D reconstruction and rendering, demonstrating significant potential. It directly generates 3D assets from real-world RGB images, enabling high-quality and visually convincing real-time rendering.

A major limitation of many 3D Gaussian-based methods is their inability to interact with varying scene lighting. Recent research efforts have focused on inferring the physical material properties of 3D Gaussians and developing relighting techniques to make them respond to changes in lighting \cite{gaoR3DG2023,liang2024gsir,bi2024gs3, guo2024prtgs}. 
However, these relighting methods have limitations: they cannot accommodate dynamic geometry or scene changes in real-time, fail to compute indirect illumination responsive to scene variations, and lack global illumination updates under complex lighting settings. There are also recent studies focusing on volumetric light transport in 3D Gaussian primitives \cite{zhou2024unified} \cite{condor2024dontsplat}, but they require heavy computation and thus are incompatible with real-time rendering.
Consequently, real-time relighting of dynamic scenes with global illumination effects under complex lighting conditions remains an open challenge. To date, no existing method has successfully satisfied all the demands.

Real-time global illumination (RTGI) plays a crucial role in delivering immersive and visually compelling virtual environments. It is indispensable in applications like gaming, virtual reality, and architectural visualization.
Over the years, RTGI techniques have been well developed, enabling high-quality illumination for mesh representations \cite{boisse2023GI10, wright2022lumen}. Often, they leverage many approaches, including lighting classification, light sampling \cite{boksansky2021manylightsGridRes}, radiance caching \cite{wright2022lumen} \cite{boisse2023GI10}, filtering and denoising \cite{lambru2021RTGIanalysis}, etc., to accelerate rendering and improve visual fidelity.
However, the challenge of RTGI for 3D Gaussian models remains largely unexplored. Due to their fundamentally different characteristics from meshes, accurately capturing and simulating indirect illumination on them in real time presents significant difficulties.

Recently, accelerated querying operations for 3D Gaussians have emerged, such as point-cloud ray tracing \cite{gaoR3DG2023,nvidia3DGSRT2024}. 
It enables the computation of visibility, shadows, reflections, and refractions, significantly enhancing rendering realism. For 3D Gaussian models, 
point cloud tracing serves as a critical foundation for advanced lighting algorithms on 3D Gaussians, making our pipeline feasible.


In this paper, we present an efficient real-time global illumination method along with a pipeline for compound scenes containing both 3D Gaussian objects and meshes.
Specifically, we propose a formulation based on surface Light Transport Equation (LTE) for 3D Gaussians, and propose fast compound stochastic ray tracing and an optimized 3D Gaussian rasterizer, solving the crucial performance issues involving 3D Gaussians in real-time global illumination. Additionally, we adapt and integrate several rendering techniques on 3D Gaussians, such as light grid, two-level radiance cache, reflection lobe tracing, and denoising, into a unified real-time global illumination pipeline for 3D Gaussians. 

To the best of our knowledge, our study is the first exploration on real-time realistic rendering of dynamic 3D Gaussian scenes with global illumination. Furthermore, our approach offers insights into the realistic rendering of 3D Gaussian models reconstructed from the real world, contributing to the enhanced visual fidelity of natural scenes.

Overall, our main contributions are as follows:
\begin{itemize}
    \item We propose a formulation based on surface LTE for light transport on 3D Gaussians, which serves as the foundation for real-time GI applications.
    \item We propose compound stochastic ray tracing and optimized rasterization techniques to address the critical performance challenges in real-time GI of 3D Gaussians, enabling realistic rendering of indirect illumination between 3D Gaussian models and between 3D Gaussians and meshes.
    \item We present a practical real-time dynamic GI pipeline for scenes of 3D Gaussians, is able to handle dynamic scenes, glossy materials, and complex illumination conditions with various light settings.
\end{itemize}

\section{Related Work}
\subsection{Real-time Global Illumination}
Achieving Real-time Global Illumination (RTGI) remains a central challenge in computer graphics. Foundational offline methods, including Path Tracing \cite{kajiya1986rendering}, Bidirectional Path Tracing \cite{veach1998robust}, Photon Mapping \cite{jensen1996pm, hachisuka2008ppm}, and their variants \cite{kaplanyan2013appm, lin2020cppm, lin2023fvcm, popov2015probabilistic, su2022spcbpt, su2025proxy}, excel at producing convincing lighting effects but are computationally expensive. Consequently, optimization strategies such as path guiding \cite{muller2017practical, ruppert2020robust, dong2023neural, huang2024online} and adaptive sampling \cite{hachisuka2008multidimensional, rousselle2011greedy, rousselle2012adaptive, salehi2022deep} have been widely adopted. Nevertheless, they also inspire optimizations on the shift to real-time rendering. These techniques are valuable for accelerating radiance cache updates when the ray budget is tight.

Bringing global illumination to real-time applications is more challenging and usually involves radical approximations. Classic methods include voxel cone tracing \cite{crassin2011interactive} and light propagation volumes \cite{kaplanyan2010lpv}. Recent advances in hardware ray tracing have led to improved techniques with radiance caching and better sampling strategies. 
Majercik et al. \cite{majercik2019DDGI} proposed a probe volume radiance cache that continuously updates and responds to changes in the scene. Meanwhile, Bitterli et al. \cite{bitterli2020ReSTIRDI} explored spatial-temporal sample reuse through resampled importance sampling with reservoirs. Later, Ouyang et al. \cite{ouyang2021ReSTIRGI} extended this sampling and reuse mechanism to global illumination, and Lin et al. \cite{daqi2022GRIS} enhanced the underlying theories. Wright et al. \cite{wright2022lumen} and Boissé et al. \cite{boisse2023GI10} developed comprehensive GI pipelines by combining various rendering, caching, and sampling techniques tailored to different types of lighting based on their characteristics, which has become the prevailing approach to achieving superior visuals. 
Recent advances in RTGI are usually capable of handling dynamic scenes and producing high-quality lighting. However, they do not support scenes with 3D Gaussian models due to the lack of a sound and easy-to-solve light transport model.

\subsection{Neural Radiance Fields (NeRF) and 3DGS}
NeRF \cite{mildenhall2020NeRF} uses an implicit differentiable multi-layer perceptron (MLP) for volume representation and reconstruction via volume rendering.
Research has focused on improving NeRF's scalability and efficiency \cite{muller2022instant,barron2021mip}, as well as decomposing lighting, extracting material properties, and enabling relighting \cite{boss2021nerd,jin2023Tensoir,bi2020NeRFRelightable}. However, NeRF-based 3D representations require time-consuming volume rendering, unavailable for real-time applications.
3DGS \cite{kerbl20233dgs} offering real-time performance by representing 3D geometry as varying Gaussian distributions, with outgoing radiance specified as spherical harmonics and rendered using the EWA splatting algorithm \cite{zwicker2002ewa} with rasterization. 3DGS has inspired numerous advancements, including improvements in forward and backward efficiency \cite{feng2024flashgs,wang2024adr}, support for large-scale scenes \cite{wang2024pygs,liu2025cityGaussian}, specialized materials \cite{liu2025mirrorGaussian}, and extended operations on 3D Gaussian datasets \cite{yu2024mip,nvidia3DGSRT2024}. However, by relying on static radiance inferred from training images, 3DGS lacks physically-based light transport, limiting its ability to handle dynamic lighting conditions or scenes.

\subsection{Relighting}
Relighting NeRF models has been widely studied, with efforts focusing on feature integration, geometric detail recovery, and lighting-aware decoding, achieving limited success \cite{bi2020NeRFRelightable, boss2021nerd,jin2023Tensoir,zeng2023nrhints}.
Recent advancements in 3D Gaussians have led to several works, such as \cite{liang2024gsir}, which proposed an inverse and relighting rendering framework, and \cite{jiang2024gaussianshader,wu2024deferredgs}, which explored traditional shading models for relighting. \citet{gaoR3DG2023} introduced inverse rendering and point cloud tracing for lighting decomposition and material optimization, while \citet{bi2024gs3} and \citet{fan2024rng} developed neural network-based relighting pipelines for 3D Gaussian models with ill-formed surfaces. \citet{guo2024prtgs} use precomputed radiance transfer for real-time relighting.
These methods rely on pre-training and offline processing, supporting only limited lighting conditions like a single environment or point light. As a result, they fail to enable real-time relighting with dynamic scenes and complex lighting.

\subsection{Shading Model and Light Transport}
Prior relighting and reconstruction methods have explored physically-based shading models for 3D Gaussian models, such as \cite{jiang2024gaussianshader,wu2024deferredgs,gaoR3DG2023,ye20243dgsreflection}.
\citet{lumaai} developed a Unreal Engine (UE) plugin integrating 3D Gaussian models into UE’s real-time rendering pipeline, but it lacks support for complex multi-bounce lighting. \citet{chen2024gi} addressed offline GI for 3D Gaussians but is limited to a narrow lighting range and treats them as solid meshes with screen-space-only light transport. \citet{guo2024prtgs} proposed a transfer equation, but stayed ambiguous on the visibility term in 3D Gaussian light transport.

There are also studies that utilize volumetric light transport for GI with 3D Gaussian primitives, such as the linear transmittance model \cite{zhou2024unified} and the exponential scattering media model \cite{condor2024dontsplat}. Both offer theoretical foundations for light transport, inverse rendering, and GI on 3D Gaussian datasets. However, they are offline methods that rely on computationally expensive volumetric light transport, making them impractical for real-time applications. Moreover, their transmittance derivations diverge from the rasterization-based approach widely used. They can not directly relight models from current inverse 3D Gaussian renderers.

\section{Our Method}
Simulating multi-bounce light transport on 3D Gaussians poses two key challenges: formulating the Light Transport Equation that enables real-time global illumination and addressing performance bottlenecks in ray tracing and rasterization. In this section, we identify the remaining obstacles to real-time global illumination for 3D Gaussian scenes and present our solutions.

\subsection{Light Transport on 3D Gaussians Model}
The LTE defines global illumination we're trying to solve. To ensure consistency with current 3D Gaussian rasterizers while maintaining real-time performance, we derive a modified surface-form LTE that is computationally more efficient than previously proposed volumetric LTEs \cite{zhou2024unified} \cite{condor2024dontsplat}.

Excluding transmissive materials, the classical LTE on surfaces \cite{kajiya1986rendering} is as:
\begin{equation}
L_o(p, w_o) =  \begin{array}{l} L_e(p, w_o) + \\ \int_{S} f(p, w_i, w_o) L_o(q, -w_i)  V(p, q)G(p, q) \cos\theta  \mathrm{d}q\end{array} \ ,
\label{eq:LTE-original}
\end{equation}
where $w_i = \| q-p \|$, and the visibility term
\begin{equation}
 V(p, q) = \left \{ \begin{array}{ll} 1 ; & \; \text{if} \; p\leftrightarrow q \; \text{is not occluded.} \\ 0 ; & \; \text{otherwise.}    \end{array} \right.
\label{eq:binary-visibility}
\end{equation}
and the geometry term $G(p, q) = \frac{-\mathbf{n_q} \cdot w_i}{|p-q|^2}$.

In 3D Gaussian rasterization, the translucency between a pair of spatial positions varies with splatting configurations. We select orthographic projection and a splatting direction $\mathbf{v}=\| q-p \|$ to derive the translucency.
\begin{equation}
T(p, q) = \prod_{i=1}^N (1 - A_{g_i, \mathbf{v}}(r)\cdot \frac{1+\mathrm{sgn}(p-p_i, \mathbf{v}) \cdot \mathrm{sgn} (q-p_i, \mathbf{v})}{2}),   
\end{equation}
where $p_i$ is the center of i-th 3D Gaussian $g_i$, and$A_{g_i, \mathbf{v}}(r)$ is the opacity of the intersection produced by intersecting the splatted i-th 3D Gaussian on direction $\mathbf{v}$ with ray $r = (p, \textbf{v})$.

Since the surfaces produced by splatting 3D Gaussians are not opaque, we replace the binary visibility term \autoref{eq:binary-visibility} with $T(p, q) \cdot A_{\mathbf{v}}(r)$ in \autoref{eq:LTE-original}, where $A_\mathbf{v}(r)$ is short for $A_{g_q, \| q-p \|}((p, \| q-p \|))$, accounting for semi-transparent surfaces. The splatted 3D Gaussian surfaces are influenced by the splatting directions, so LTE is expressed in terms of integration over spherical angles, giving: 
\begin{equation}
L_o(p, w_o) =  L_e(p, w_o) + \int_{H^2} f(p, w_i, w_o) L_i(p, w_i)  \cos\theta \mathrm{d}w_i \ ,
\label{eq:LTE-modified}
\end{equation}
The scattering function $f(p, w_i, w_o)$ and surface normal at $p$ should be derived from the material attributes on the hit 3D Gaussian at $p$.
For direction $w_i$, there can be multiple splatted surfaces with non-zero contribution to incoming radiance $L_i(p, w_i)$, so we have to enumerate them as:
\begin{equation}
L_i(p, w_i) = \sum_{i=1}^n T(p, q_i) A_\mathbf{v}(q_i) L_o(q_i, w_i),
\label{eq:enumerate-intersection}
\end{equation}
where $q_1, ..., q_n$ are intersections of ray $(p, w_i)$ with 3D Gaussians splatted in direction $\mathbf{v}$. 
Along the ray traced,  weighted contributions of the intersections should be added up to produce the incident radiance $L_i(o, \textbf{v})$.

For now, we have derived the LTE for pure 3D Gaussian models. To incorporate traditional mesh-based geometries $\mathrm{G}$, we simply set $A_\mathbf{v}((p, \mathbf{v}))=1$ for any $\mathbf{v}$ and all $p\in \mathrm{G}$. 
With the problem formally defined, we now focus on developing an algorithm to approximate the solution to this equation.

\subsection{Overall Pipeline}

\begin{figure*}
    \centering
    \includegraphics[width=0.95\linewidth, trim={0 0 0 0}]{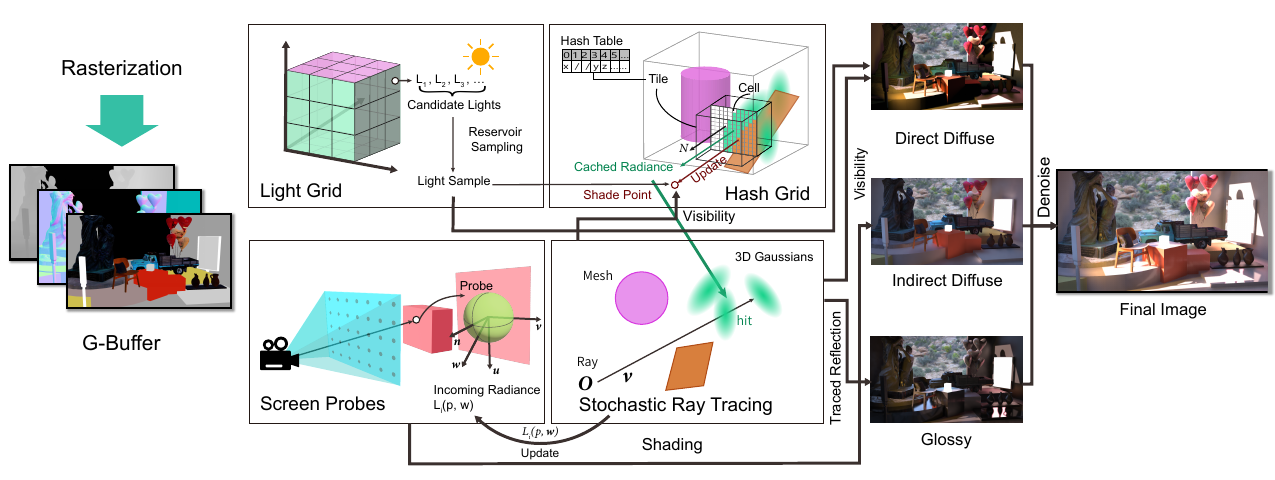}
    \caption{Our RTGI pipeline for 3D Gaussian models. The pipeline uses light sampling from light grid and shadow ray-tracing for direct diffuse lighting. The two-level-cache including the hash grid and screen probes maintains indirect diffuse lighting. Shading ray-tracing that queries intersection properties is used to update the cache and render glossy reflectance. }
    \label{fig:pipeline-structure}
\end{figure*}

In the context of real-time GI, we adopt certain assumptions to simplify the computation of the full LTE, as follows:

\begin{itemize}
    \item \textbf{Simplified material model:} For the primary vertices along light paths, we employ a combination of the GGX and Lambertian material models. For the remaining vertices we assume a purely Lambertian model, finding a balance between accuracy and computational efficiency.
    \item \textbf{Clipped light path:} Light paths that exceed two bounces and cannot be \textit{recursively accumulated on the film} are discarded, which will be discussed in \autoref{subsec:twocache}.
\end{itemize}

Our approach follows current RTGI pipelines using ray tracing. 
Final illumination is classified into four components based on material properties of primary vertices: emission, diffuse (direct and indirect), and glossy reflectance, with each component rendered separately using specialized algorithms. 

In our pipeline (see \autoref{fig:pipeline-structure}), we first rasterize 3D Gaussians and meshes to generate the G-buffer, including emission maps. Then, direct diffuse lighting is computed by sampling lights per pixel from a dynamically constructed light grid, tracing shadow rays for visibility, and spatio-temporally filtering the results. For indirect diffuse lighting, we use a two-level radiance cache: the primary cache with screen light probes \cite{wright2022lumen} and the secondary cache with a hash grid cache \cite{boisse2023GI10}, which are temporally reused and progressively updated to reduce noise. Spherical harmonic coefficients from the screen light probes are interpolated and multiplied with the material BRDF for shading. Glossy reflectance is computed by tracing rays in the reflection lobe of glossy surfaces, followed by spatio-temporal filtering and accumulation, using a split-sum approximation for shading. Finally, tone mapping is applied to the combined radiance to produce the final result. This pipeline effectively handles various lighting types, optimizing both performance and visual quality in dynamic scenes.
More technical details will be discussed in \cref{sec:implementation}.

Specifically, 3D Gaussian models and meshes are rasterized altogether to get a G-buffer, and the geometry reconstructed from the G-buffer is the weighted average of all surfaces enumerated in \autoref{eq:enumerate-intersection}, expressed as: 
\begin{equation}
\overline{q} = \sum_{i=1}^{n}T(p, q_i)A_{\mathbf{v}}(q_i)\cdot q_i \ .
\label{eq:average-surface}
\end{equation}
Using $L_i(p, w_i) \approx (\sum_{i=1}^{n}T(p, q_i)A_{\mathbf{v}}(q_i)) L_o(\overline{q}, w_i) $, we further apply a split-sum approximation \cite{karis2013real} for the first bounce of light paths, reducing the shading operation from multiple evaluations to just once per screen pixel.

Despite the assumptions and simplifications outlined earlier, prior ray-tracing-based RTGI methods remain a performance-bottleneck due to the high computational cost of the latest 3D Gaussian ray-tracing methods \cite{nvidia3DGSRT2024, gaoR3DG2023} and rasterizers. To address this issue, we propose a more efficient compound stochastic ray-tracing algorithm and an optimized forward-only rasterizer.

\subsection{Compound Stochastic Ray-Tracing}

\begin{figure}
    \centering
    \includegraphics[width=\linewidth, trim={0 3cm 0 2cm}]{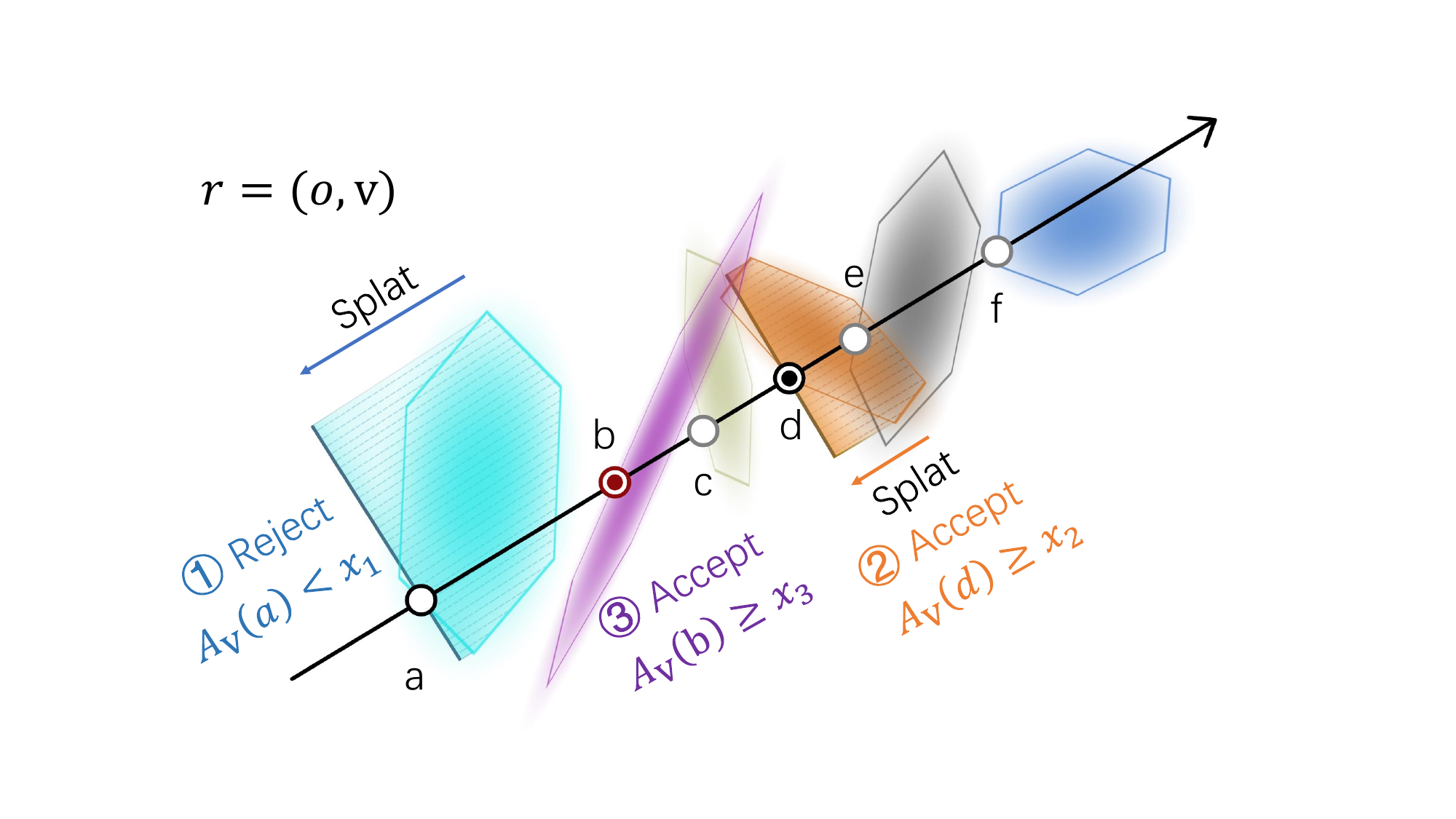}
    \caption{Illustration of a shading ray traced in our algorithm. 
    Hardware ray tracing does not guarantee a specific \textit{anyhit} invocation order along the ray; we illustrate one possible permutation of intersections.
    First, the GPU reports an intersection with the proxy geometry at \textit{a}, but the hit is rejected as its opacity \( A_\mathbf{v}(a) \) is lower than the random threshold \( x_1 \). Then, the GPU reports intersection \textit{d}. The orange Gaussian passes the opacity test, culling subsequent intersections (\textit{e}, \textit{f}). 
    Finally, \textit{b} is accepted, culling \textit{c}. As the closest hit, \textit{b}'s features are returned as the trace result.}
    \label{fig:stochastic-rt-proc}
\end{figure}

\begin{figure}
    \centering
    \begin{tabular}{@{}c@{}c@{}}
        \includegraphics[width=0.6\linewidth]{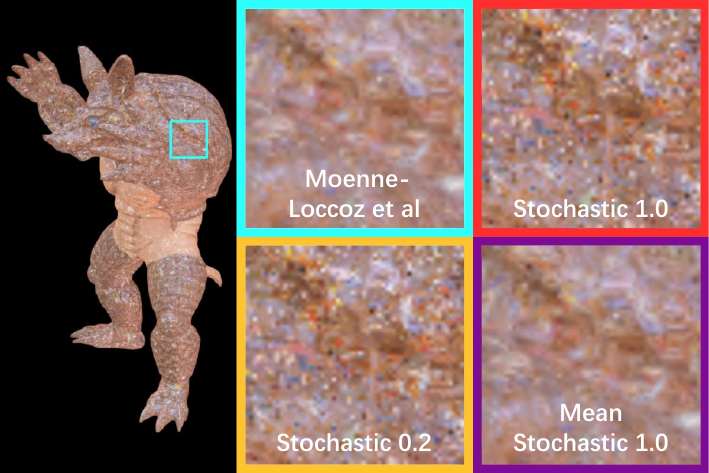} &\
        \includegraphics[width=0.25\linewidth]{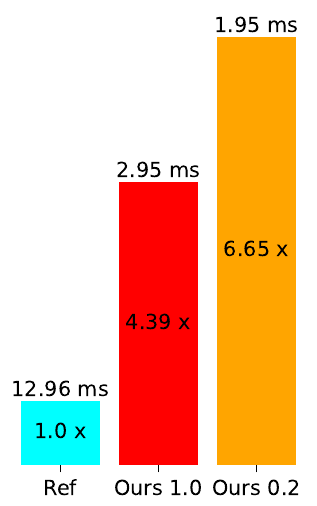} \\
    \end{tabular}
    \caption{Ray-traced hit feature (albedo) from \citet{nvidia3DGSRT2024} and our stochastic ray tracing, respectively. $A_\mathbf{v}(r)$ in stochastic ray tracing is replaced with \textit{Gaussian Max-response} for consistency. The initial random value is scaled by 1 or 0.2. We visualize the mean of the unscaled stochastic ray-trace result across multiple frames, which can be identical to \citet{nvidia3DGSRT2024}. We also show the performance gain in ray throughput.}
    \vspace{-0.1in}
    \label{fig:stochastic-rt}
\end{figure}

A crucial observation is that only unbiased estimators rather than actual values for ray-trace results are needed. Thus, we introduce a stochastic process to ray-tracing, trading enormous performance with endurable noise.

Replacing the \textit{Gaussian Max-response} with $A_\mathbf{v}(r)$ as particle response, we adopt a modified version of the ray-tracing algorithm from \cite{nvidia3DGSRT2024} as our reference. Rather than computing the exact value of \autoref{eq:enumerate-intersection}  by enumerating all possible intersections along a ray, we propose a theoretically unbiased estimator using stochastic ray tracing.



We use stretched polyhedral \textit{proxy geometries} to bound Gaussians and build an acceleration structure for hardware ray-tracing, consistent with the reference algorithm. While intersection orders may differ from classical rasterization, empirical results show that the introduced error is negligible for shadow rays, and endurable for shading rays thank to the denoising in our two-level-cache.

\textbf{Shadow Rays} determine whether a ray segment intersects with the scene. A ray $r = (\mathbf{o}, \mathbf{v})$ with origin $\mathbf{o}$ and direction $\mathbf{v}$ is dispatched. When an \textit{anyhit} shader is invoked on a Gaussian proxy geometry with unspecified order, the hit's particle response $A_\mathbf{v}(r)$ is evaluated and compared to a random number \textit{x} between 0 and 1. If the hit response $A_\mathbf{v}(r)$ is less than \textit{x}, tracing continues; otherwise, the intersection is confirmed. This process enables early termination on high-opacity Gaussian surfaces, reducing computational costs.


\noindent \textbf{Property:} $E(b) = 1 - T(r)$, where $b$ is the expectation for the shadow ray tracing result (1 for occluded), and $T(r)$ is the ray translucency to which the reference algorithm evaluates. 

We provide the proof in the supplementary. The above property makes our stochastic shadow ray tracing a competent estimator for the fraction of non-occluded radiance transported along a ray segment. It is useful in Monte-Carlo-based direct lighting algorithm.

\textbf{Shading Rays} are the rays querying intersection properties for shading purposes. Different from shadow rays, we enable the \textit{closesthit} shader. Thus, the closest real hit is reported as the authentic hit. Features like material properties are extracted from the hit 3D Gaussian as the tracing result. We show an example of the tracing process in \autoref{fig:stochastic-rt-proc}. 

\noindent \textbf{Property:} The expectation for the ray-traced hit's features equals the ray-traced result from the reference algorithm.
\begin{proof}
In essence, our algorithm is equivalent to the following process in terms of tracing outcomes:
\begin{itemize}
    \item Gather all hit Gaussians along the ray.
    \item Accept each hit by its evaluated opacity $A_{\textbf{v}, g}(r)$ as a probability. Otherwise, reject the hit.
    \item Return the feature of the closest hit among all accepted hits.
\end{itemize}
Thus, for a hit $x_i$ to end up being accepted and selected as the final hit $y$ for the result, the probability is
\begin{equation}
P(y=x_i) = A_{g_i, \mathbf{v}}(r) \cdot \prod_{j=1}^{i-1} (1 - A_{g_j, \mathbf{v}}(r)),
\label{eq:hit-probability}
\end{equation}
where $x_1, ..., x_{i-1}$ are the hits closer to the ray origin than $x_i$ in near-to-far order, and $g_1, ..., g_i$ are the corresponding 3D Gaussians.
Note that $P(y=x_i) = A_{g_i, \mathbf{v}}(r) * T_{i-1}$ where $T_{i-1}$ is the accumulated transparency till $(i-1)$-th Gaussian. 

Enumerating all hits, the expectation of the feature $f$ returned from the above process is
\begin{equation}
E(y) = \Sigma_{i=1}^n P(y=x_i) f_i = \Sigma_{i=1}^n A_{g_i, \mathbf{v}}(r)\cdot T_{i-1} f_i \ , 
\label{eq:hit-expectation}
\end{equation}
where $f_i$ is the feature value of the $i$-th Gaussian. The expectation is exactly the rendering target of the reference algorithm.
\end{proof}

By biasing accept probabilities, we can further accelerate the tracing process at the expense of unbiasedness. However, we find this trade-off worthwhile, given the significant performance gains. To balance bias and efficiency, we scale the initial random numbers in our tracing algorithm within a range of 0 to 1. \autoref{fig:stochastic-rt} compares our results with the baseline \cite{nvidia3DGSRT2024}, and illustrates the impact of the scaling factor. We observe a significant performance gain at the cost of introducing noise. Meanwhile, the per-pixel mean of our ray-trace results is identical to the baseline.

\textbf{Estimating Incoming Radiance} using stochastic ray tracing is done by simply averaging the shading results of multiple shading ray hits of the same ray. Assume that we have an estimator of the outgoing radiance $\mathrm{E}(\hat{L_o}(q_i)) = L_o(q_i)$. By picking a random intersection $q_{j}$ enumerated in \autoref{eq:enumerate-intersection}, a 1-sample Monte-Carlo estimation for $L_i(p, w_i)$ can be written as
\begin{equation}
\hat{L_i} = \frac{T(p, q_j)A_\mathbf{v}(q_j)\hat{L_o}(q_{j})}{P(q_{j})}. 
\end{equation}
From \autoref{eq:hit-probability}, we found 
\begin{equation}
P(q_j) = A_{g_m,\mathbf{v}}(r) \cdot \prod_{i=1}^{m-1} (1 - A_{g_i, \mathbf{v}}(r)),    
\end{equation}
with $g_1, ..., g_m$ as the 3D Gaussians hit along the ray by ascending order and $g_m$ the Gaussian producing hit $q_j$. The above can simply be written as $P(q_j) = T(p, q_j)A_\mathbf{v}(q_j)$. Substituting it into the original estimation, we cancel out the terms in numerator and denominator, producing $\hat{L_i} = \hat{L_o}(q_{j})$ where $q_j$ is a hit our unbiased stochastic ray-tracing algorithm produced. Thus, for the n-sample Monte-Carlo estimator, we can simply average the results from shading the hits returned by our tracing algorithm to get an unbiased estimator of $L_i(p, w_i)$. This becomes useful when updating the screen probe cache later described in \autoref{subsec:twocache}.



\textbf{Compound Tracing:} Following Lumen \cite{wright2022lumen}, rays start from directly visible surfaces using cheap screen-space tracing on the Hi-Z buffer, stopping at intersections or occlusions. The remaining segment is traced using stochastic ray-tracing. 
Screen-space tracing hits are sometimes biased evaluating \autoref{eq:enumerate-intersection}. We strip uncertain screen-space hits, making it more of an optimization to fast-forward away from dense geometries near ray origins.


\subsection{Optimized 3D Gaussian Rasterization}

Recent works accelerate 3D Gaussian rasterization via SM occupancy, memory optimization, and Gaussian culling \cite{feng2024flashgs, wang2024adr}. Unlike software rasterization compromising for differentiability, our forward-only renderer leverages hardware rasterization for efficiency.

\begin{figure}[t]
    \centering
    \includegraphics[width=0.95\linewidth]{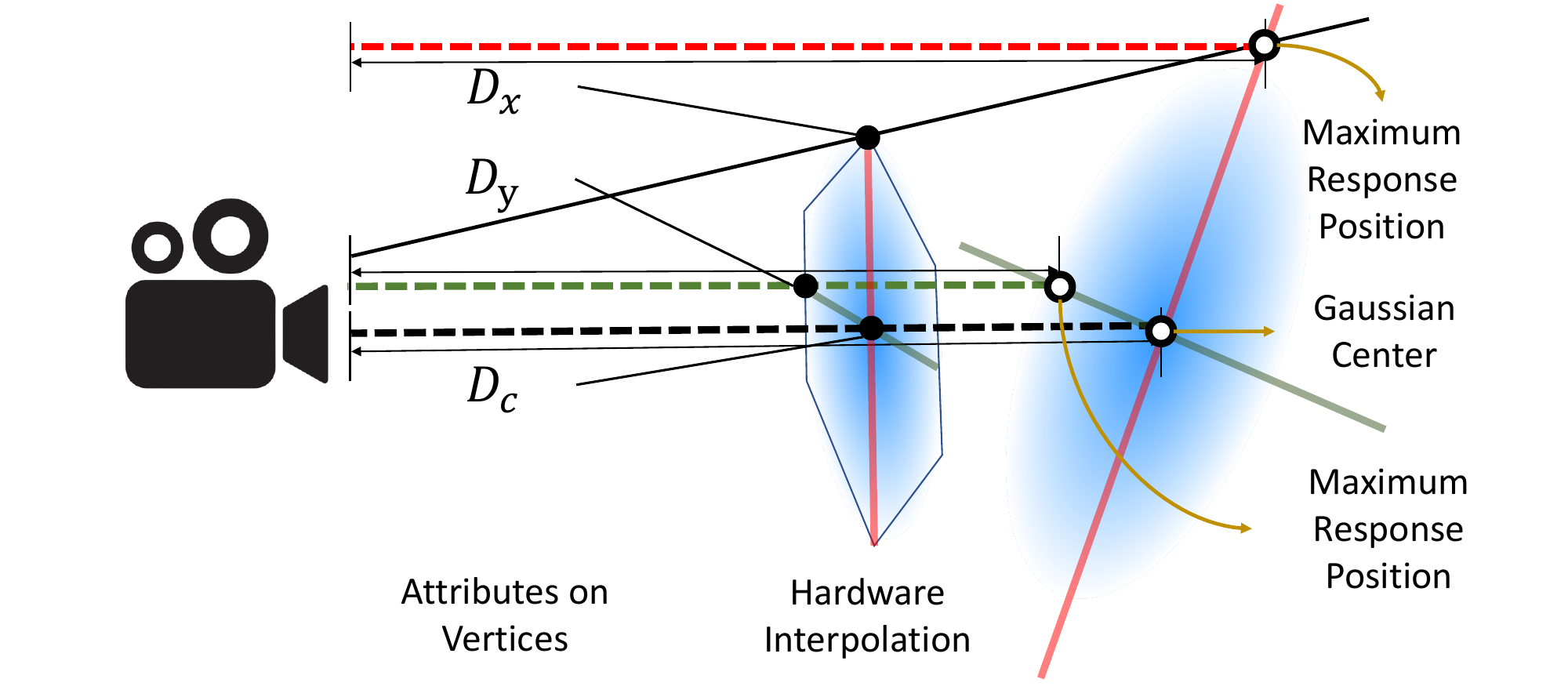}
    \vspace{-0.1in}
    \caption{At each frame, 2D hexagons are generated from 3D Gaussians for hardware rasterization, aligned with their projected distributions. Depth values (\(D_x, D_y, D_c\)) are extracted from the 3D \textit{Gaussian Max-response} positions along camera rays and stored as vertex attributes for hardware interpolation.}
    \vspace{-0.1in}
    \label{fig:hexagon-spawnning}
\end{figure}

\begin{figure}[t]
    \centering
    \begin{tabular}{cc}
    \includegraphics[width=0.45\linewidth]{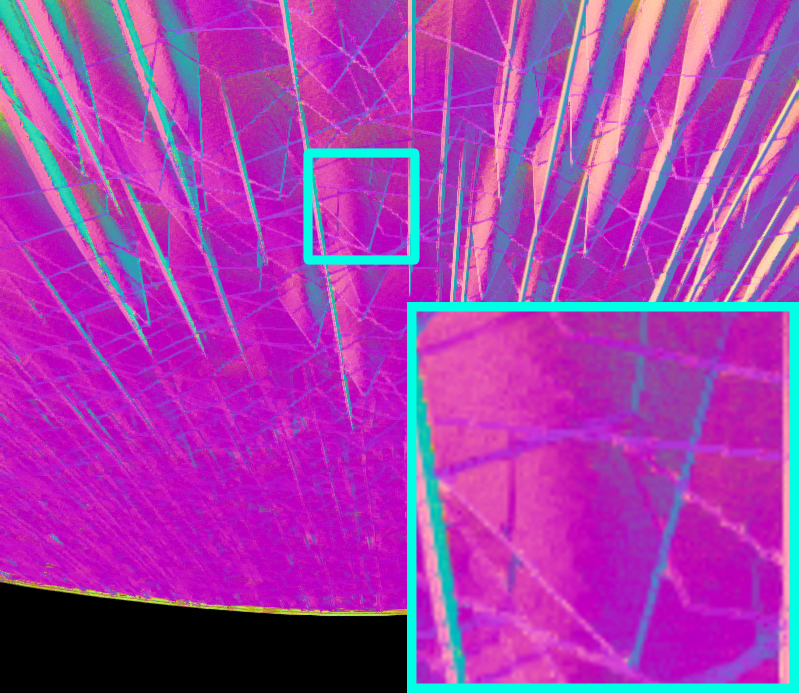} &
    \includegraphics[width=0.45\linewidth]{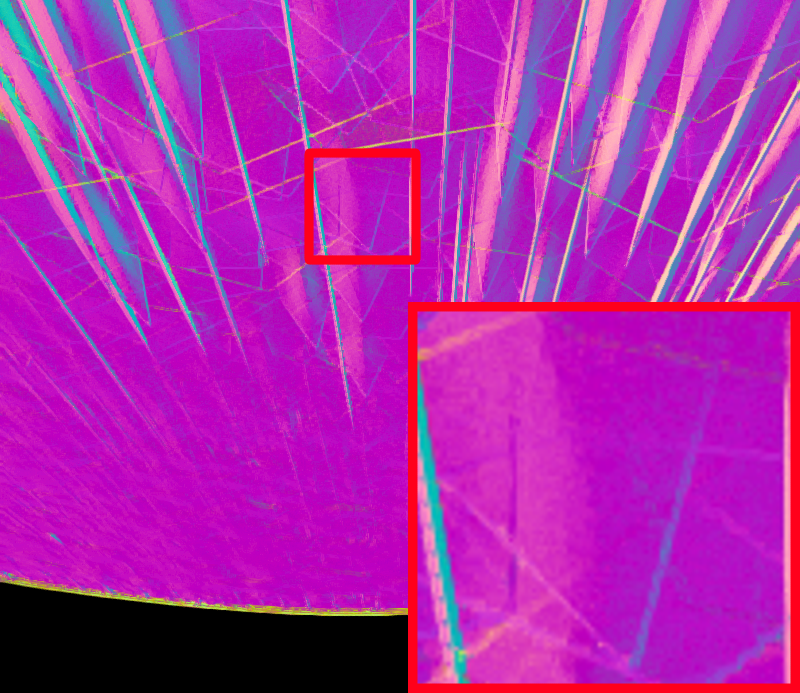}
    \end{tabular}
    \vspace{-0.4cm}
    \caption{Reconstructed normals of the flat underside of a 3D Gaussian model from depths rendered with constant Gaussian depths (left) and our approach (right). We mitigate the artifacts at the edges of hexagons, by using linear gradients for depths on each hexagon instead of constants.}
    \label{fig:depth-recon}
    \vspace{-0.5cm}
\end{figure}

3D Gaussians in the view-frustum are squashed into 2-dimensional hexagons, whose orientations and scales are determined by the eigenvalues and eigenvectors of the projected Gaussian's 2-dimensional covariance. Hexagons are hardware-rasterized in a far-to-near order.
To compensate for the edge cutoffs of the hexagon, instead of using a constant depth for all pixels of a Gaussian, we assume the depth of a 3D Gaussian projection on the camera film is a linear gradient approximating the \textit{Gaussian Max-response} plane from the camera viewpoint. \textit{Gaussian Max-response} distance from \citet{nvidia3DGSRT2024} is used as depths at the edges of the hexagons, as described in \autoref{fig:hexagon-spawnning}. 
The depth values are later hardware interpolated and blended into the final G-buffer. 
We empirically find the biased trick helpful.
\autoref{fig:depth-recon} shows the improvement by displaying the reconstructed normals from depth gradients. Our approach provides smoother surfaces.

To correct rare cases of normal inaccuracy, our rasterizer recovers \textit{fallback normals} with depth gradient. A trivial heuristic is employed to tell and replace bad rasterized normals with fallback normals.


\section{Technical Detail and Implementation}
\label{sec:implementation}

\subsection{G-Buffer}
Hardware rasterization generates the G-buffer. First, meshes are rasterized to create a depth texture. Then, 3D Gaussians undergo depth testing against the mesh depth, with blending enabled to accumulate fragments into a separate G-buffer. The two G-buffers are then merged by opacity, reconstructing the geometric and material properties of directly visible weighted-mean surfaces \( \overline{q} \). Emission is also rendered during rasterization.

\subsection{Direct Illumination}
Direct lighting is rendered with a simple Monte-Carlo method based on light sampling and ray tracing.
For each frame, lights are injected into a cascaded light grid centering at the camera with A-Res algorithm \citet{efraimidis_weighted_2006}, with weights from a heuristic estimating their contribution to grid geometries. 
Later, lights are sampled at a per-pixel level. We find the lights within the same grid cell, spawn one candidate sample per light, and sample from them with weights equal to their contributions to the pixel. 
Finally, one shadow ray for the final light sample is traced to estimate the fraction of radiance arriving at the pixel, multiplying into the overall estimator.
A spatio-temporal filter is applied to suppress the noise from 1-sample Monte-Carlo for stable diffuse direct illumination.

\subsection{Two-level Radiance Cache and Indirect Diffuse}
\label{subsec:twocache}
The two-level radiance cache is a combination of the adopted screen probe cache \cite{wright2022lumen} as the primary cache, and a modified version of hash grid cache \cite{boisse2023GI10} as the secondary cache. The cache suppresses noise produced by stochastic ray-tracing and produces indirect illumination. 

\textbf{Screen Probes} act as the primary radiance cache, storing incident radiance (excluding emission) at primary vertices. They are placed in a grid on the camera film and snapped to scene surfaces using G-buffer depth. Each probe captures hemispherical incoming radiance, with importance-sampled shading rays traced per frame for update. Probes are spatially filtered, temporally reused, and interpolated via spherical harmonics for screen pixel radiance distribution. Final shading results undergo temporal denoising for stability.


For pixels with no valid adjacent probes, they fallback to reuse lighting from nearby similar pixels.


\textbf{Hash Grid Cache} is the secondary radiance cache for our pipeline. World-space outgoing radiance queries are identified as 2-component \textit{key vectors} (position and view direction), quantized, hashed, and mapped to tiles in a hash table. Each tile is further subdivided and mapped to 2D cells based on the view direction’s major axis as shown in \autoref{fig:pipeline-structure}, optimizing storage and enabling spatial filtering.
Each frame, the shading rays allocated for screen probe update produce hits in the scene. The pipeline evaluates direct lighting for the hits using 1-sample Monte-Carlo, and accumulates results into cache cells. Quantization of \textit{key vectors} naturally handles radiance caching on ambiguous 3D Gaussian surfaces. 

\begin{figure*}
    \centering
    \includegraphics[width=0.9\linewidth]{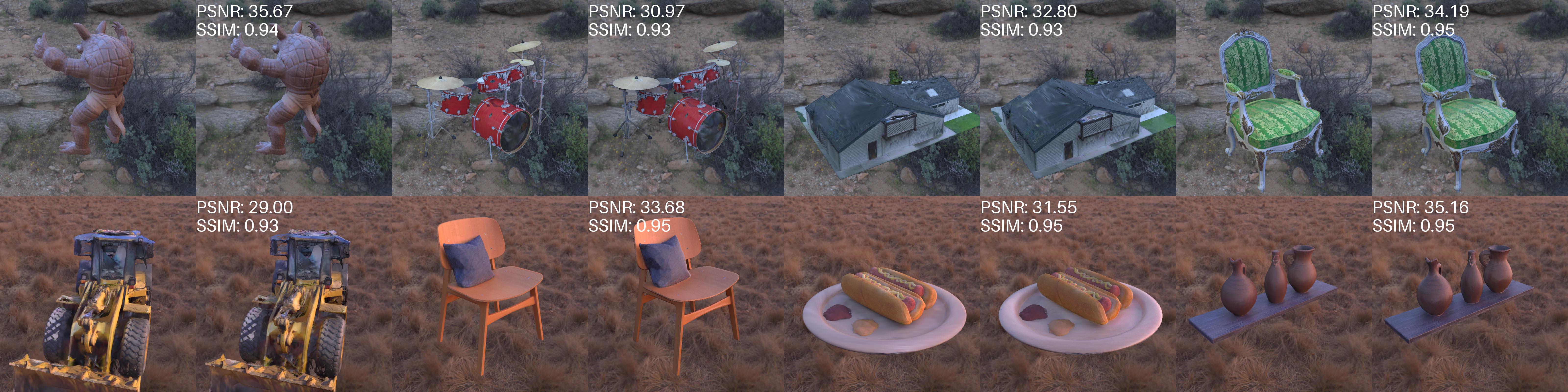}
    \vspace{-0.2cm}
    \caption{Comparison between R3DG \cite{gaoR3DG2023} and ours on various benchmarks, under environment lights. Each pair shows R3DG’s offline relighting result (left) and our real-time rendering (right). Our method achieves an average PSNR of 32.87 and SSIM of 0.94, maintaining consistency with R3DG.}
    \label{fig:r3dg-consistency}
\end{figure*}

\textbf{Indirect Diffuse Lighting:} The pipeline resolves radiance from the previous frame's film for outgoing radiance queries, and fallbacks to the secondary cache upon failure. Thus, only light paths that are no longer than 2 bounces or can \textit{recursively accumulate on the camera film} are rendered. Thus, we spare efforts to maintain indirect lighting for the secondary cache. 




\subsection{Glossy Reflectance}
The pipeline renders a ray-traced detailed reflection texture for surfaces with low roughness by sampling the glossy lobe in 1/2 the resolution, with a spatial filter for denoising. For surfaces with higher roughnesses, a coarse reflection is approximated via sampling diffuse direct lighting and adjacent screen probe texels in the reflection direction. Based on surface roughness, the detailed and coarse reflection is blended and supplied to the split-sum approximation \cite{karis2013real} for glossy rendering after denoising. 

\section{Experiment and Evaluation}


We conducted all the tests on a workstation equipped with RTX3090 GPU (24G VRAM). 
We test the consistency of our approach with the baseline. Moreover, we validate our approach through a series of experiments (dynamic scenes with diverse light settings), with all results captured from our renderer running at more than 40 fps.

\subsection{Consistency Validation}
We use the training method from Relightable 3D Gaussians (R3DG) \cite{gaoR3DG2023} to produce Gaussian models. By using the same set of optimized Gaussian models and 384 samples for R3DG as they suggested in their code base, we compare the rendering results produced by R3DG and our approach. In the single 3D Gaussian model relighting results under environment lights presented in \autoref{fig:r3dg-consistency}, we achieved consistency with R3DG. While minor artifacts occur around small occlusions and reflections due to the screen probe cache’s spatial granularity and the split-sum approximation for glossy rendering, our approach maintains high visual quality comparable to R3DG, with high PSNR and SSIM values.

R3DG only supports environment light in relighting. When relighting grouped 3D Gaussian models, inter-model multi-bounce global illumination effects become noticeable. Unlike R3DG, which cannot render these complex lighting effects, our approach fully supports them. As all subsequent experiments involve multiple models and complex light settings beyond environment light, comparisons with R3DG are no longer included.


\subsection{Various Evaluations}

\begin{figure*}
    \centering
    \begin{tabular}{c@{}c@{}c@{}c@{}c@{}c}
         \includegraphics[width=0.149999994\linewidth]{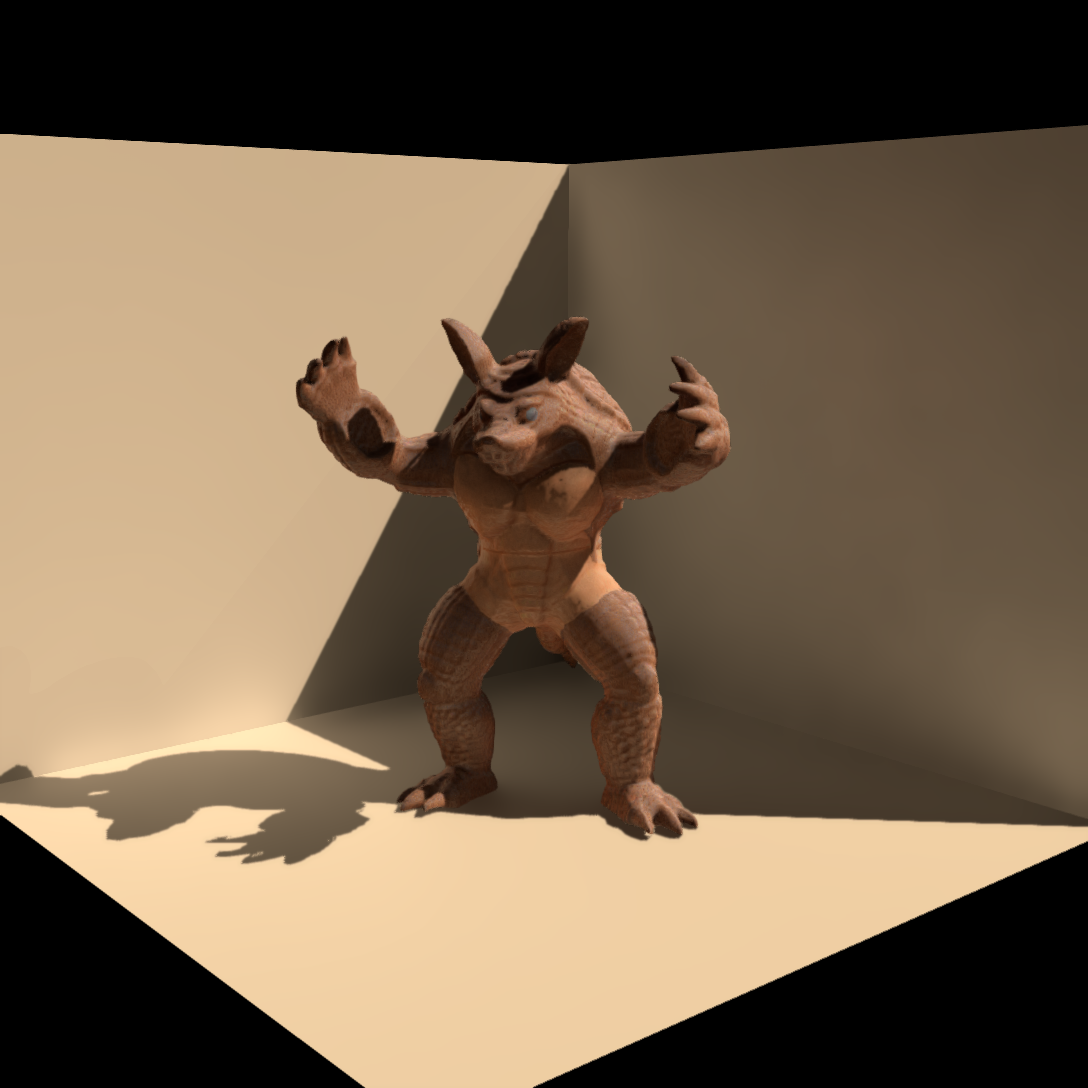}&
         \includegraphics[width=0.149999994\linewidth]{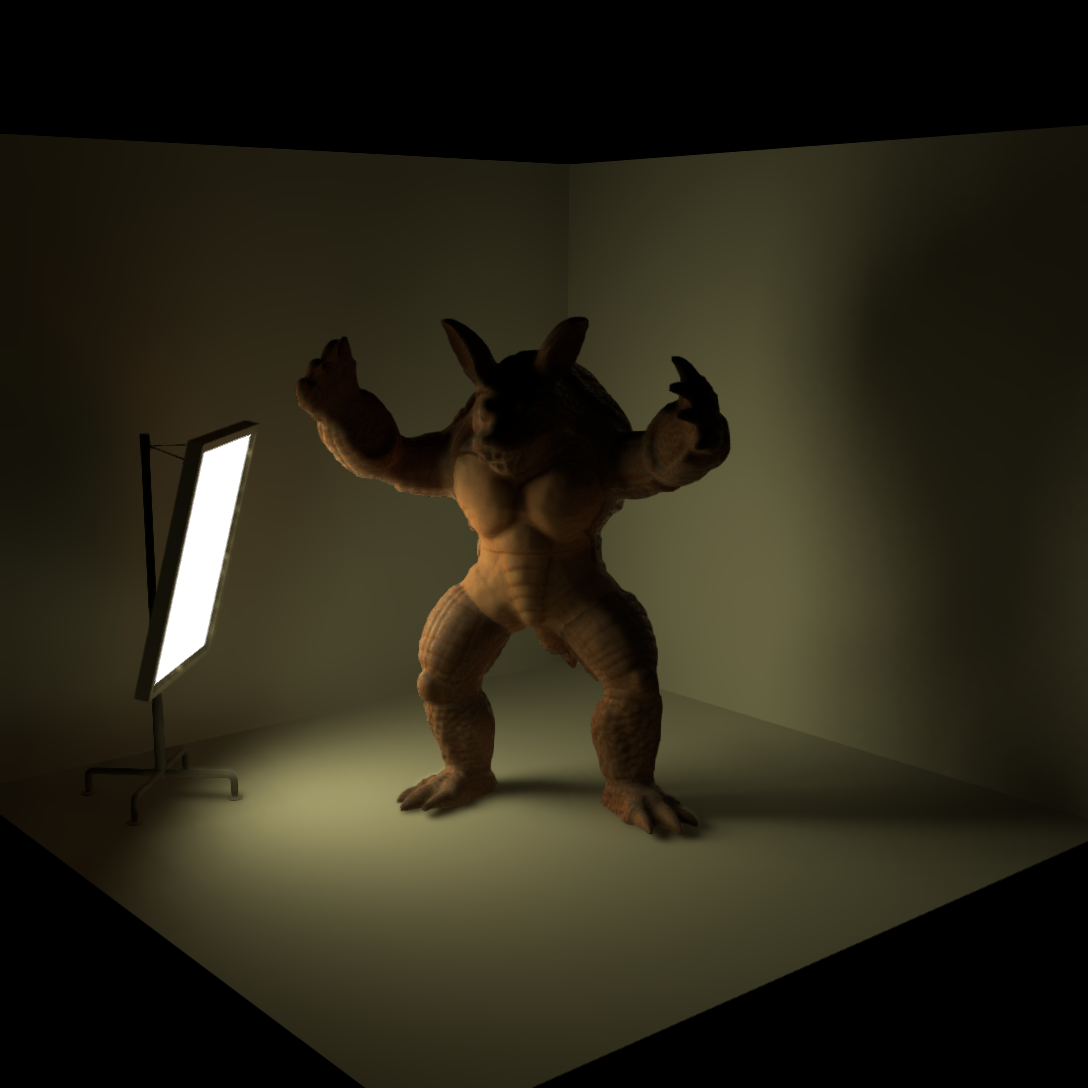}&
         \includegraphics[width=0.149999994\linewidth]{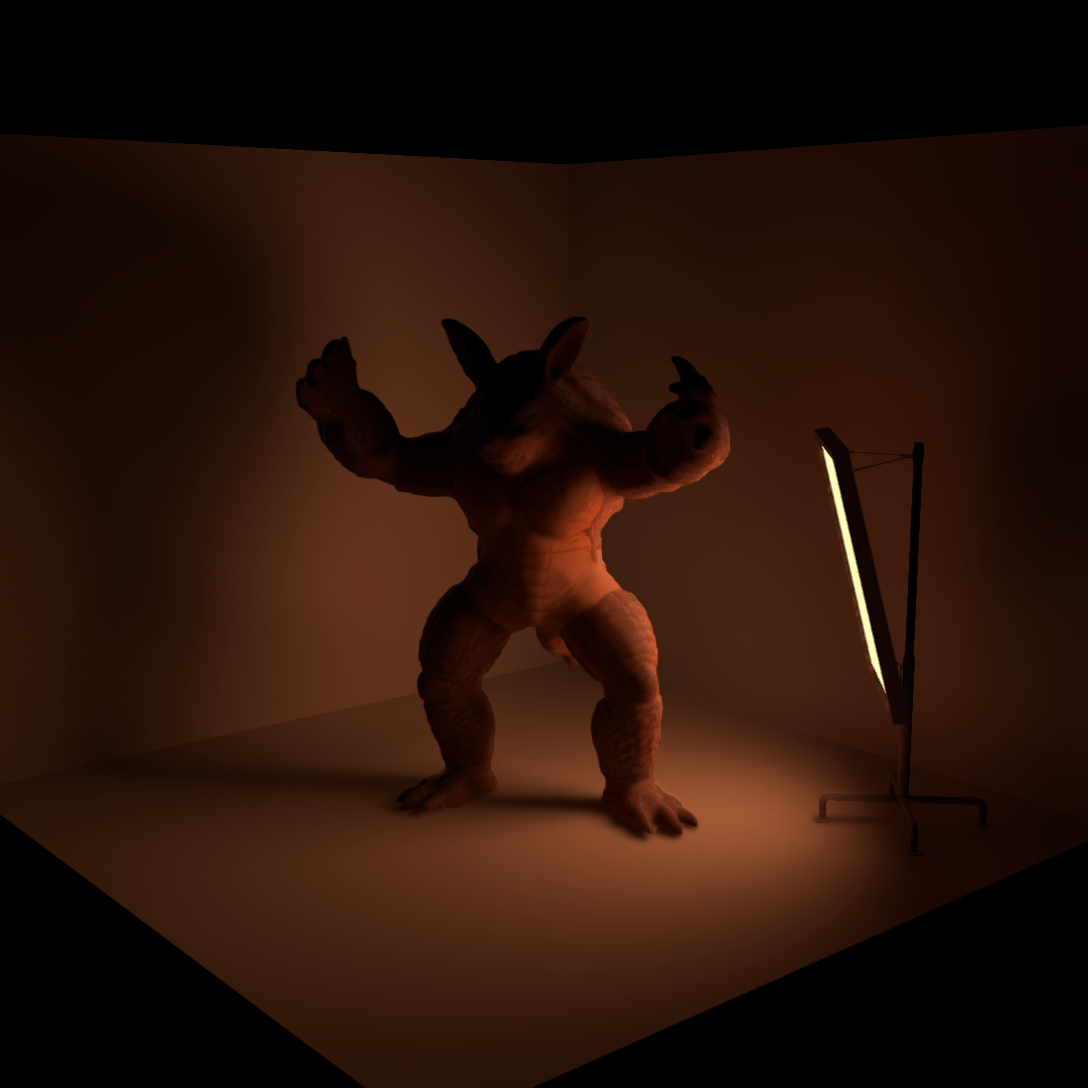}&
         \includegraphics[width=0.149999994\linewidth]{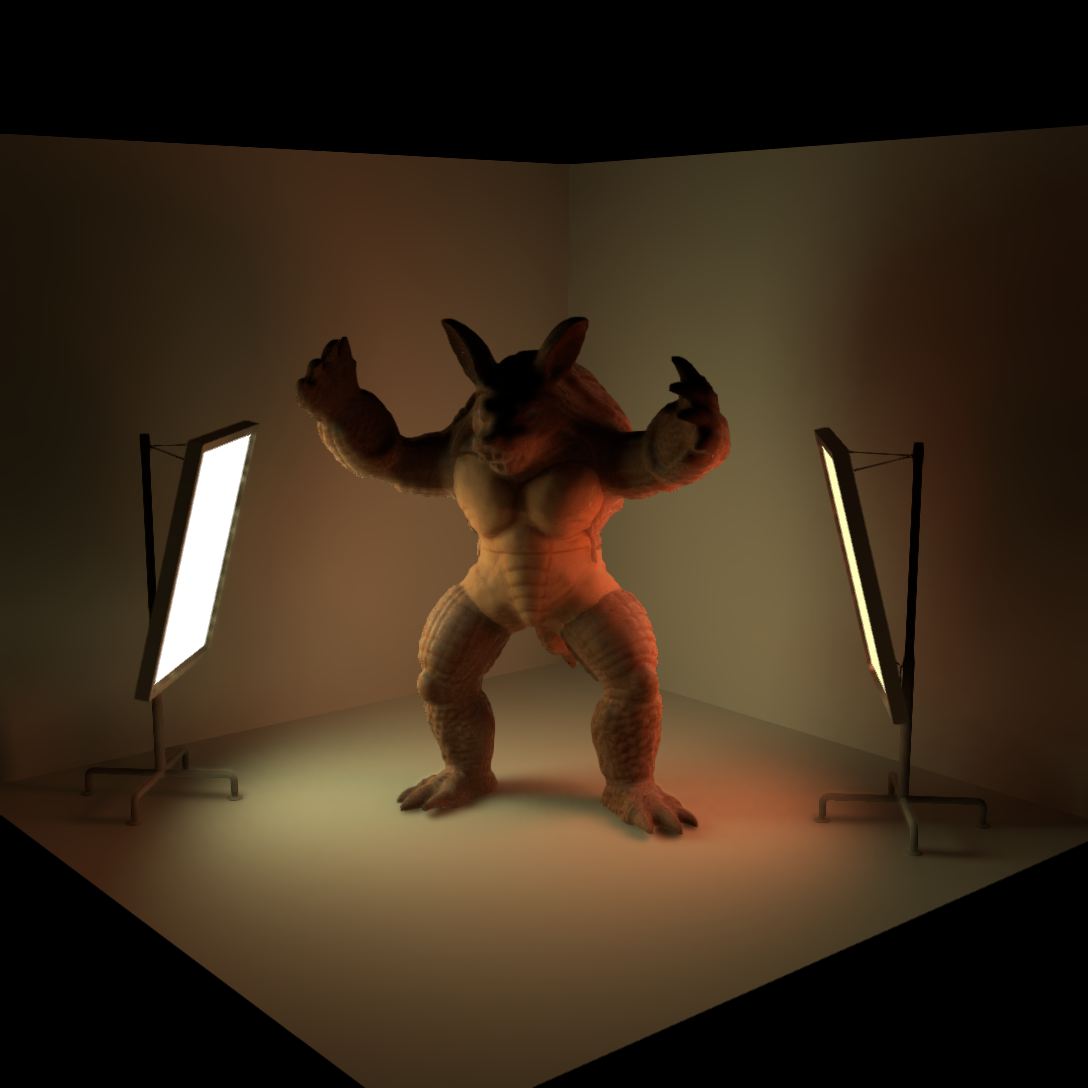}&
         \includegraphics[width=0.149999994\linewidth]{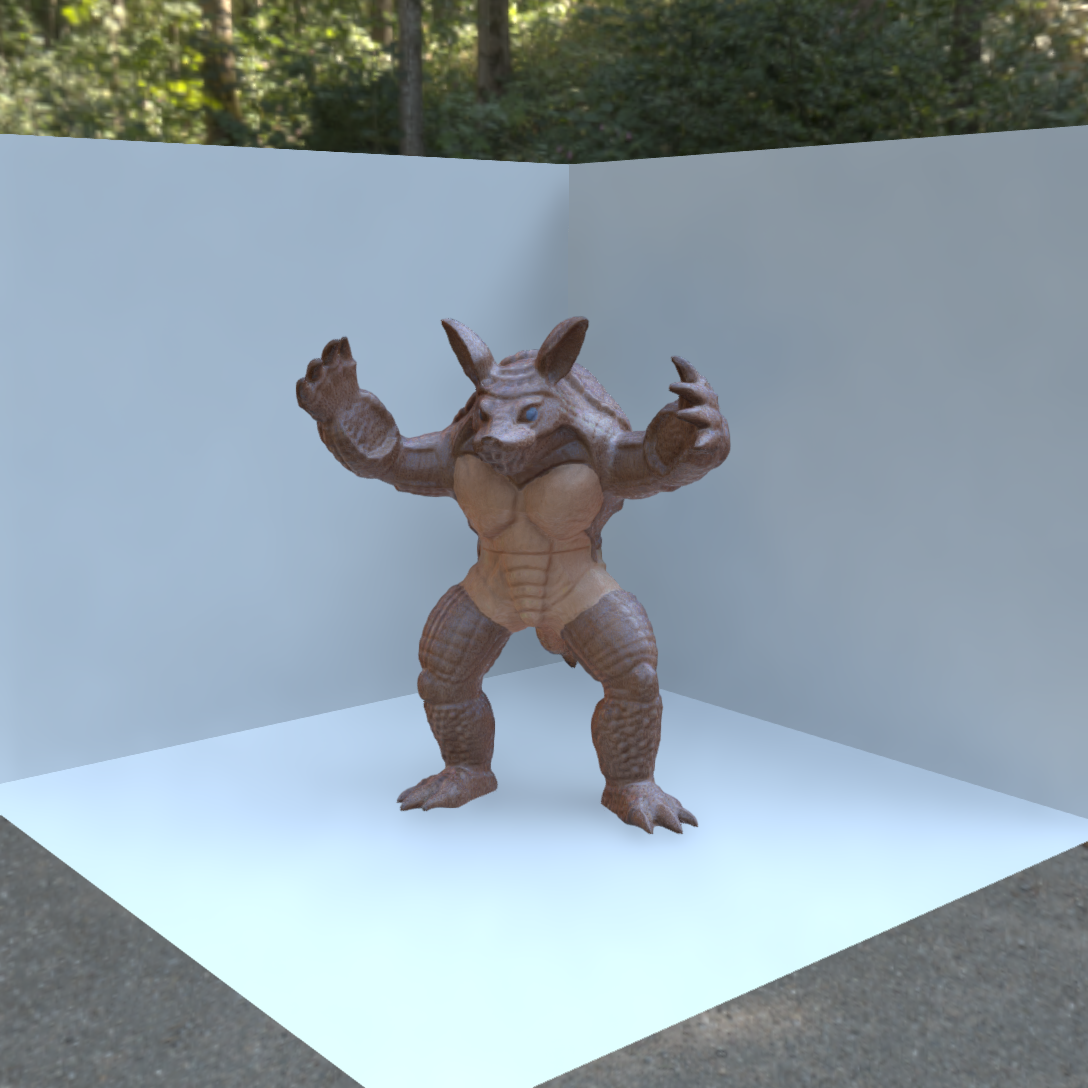}&
         \includegraphics[width=0.149999994\linewidth]{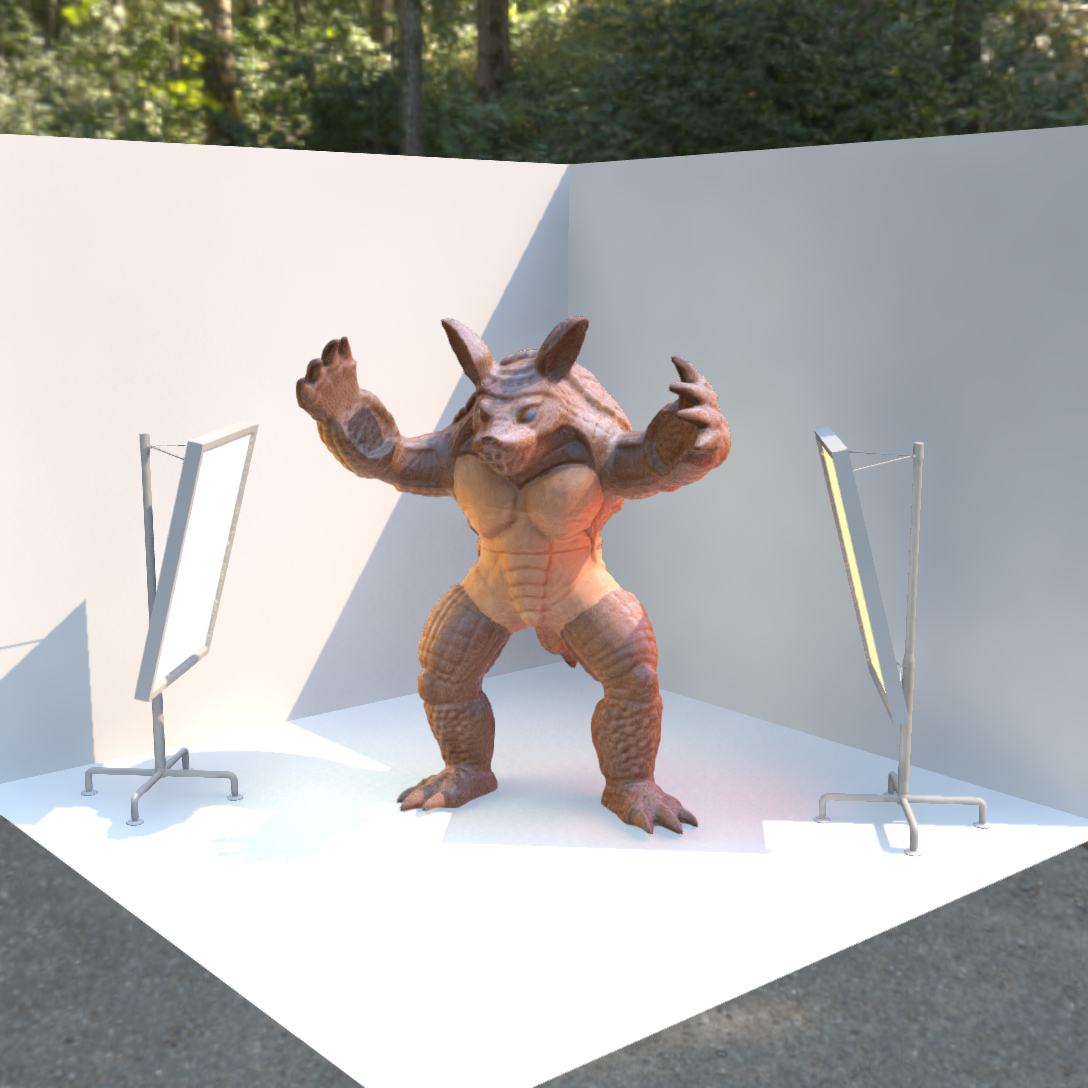}
         \\
        Directional Light & \multicolumn{3}{c}{Area Lights} & Environment Light & All
    \end{tabular}
    \begin{tabular}{c@{}c@{}c}
         \includegraphics[width=0.2999997\linewidth]{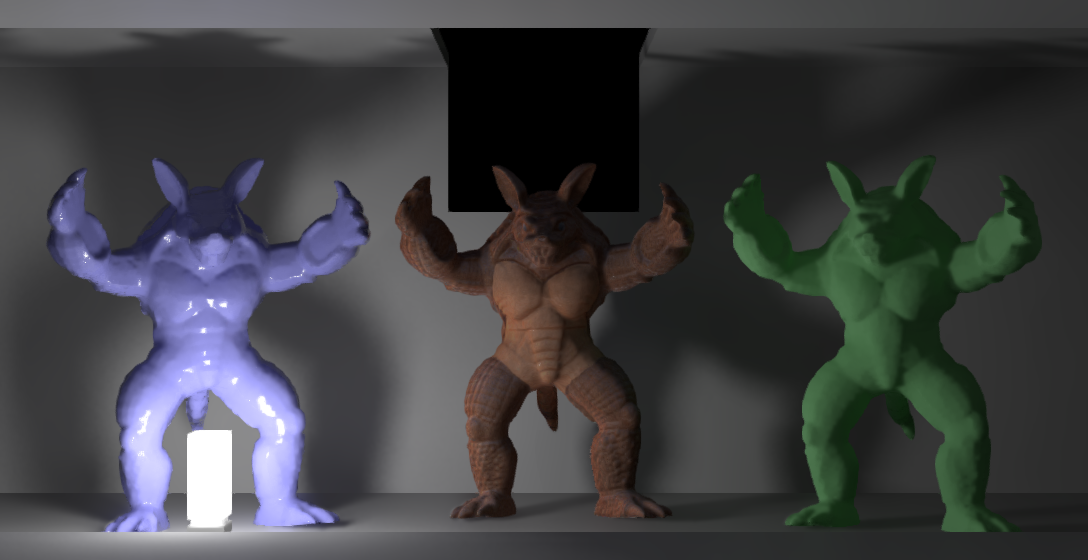}&
         \includegraphics[width=0.2999997\linewidth]{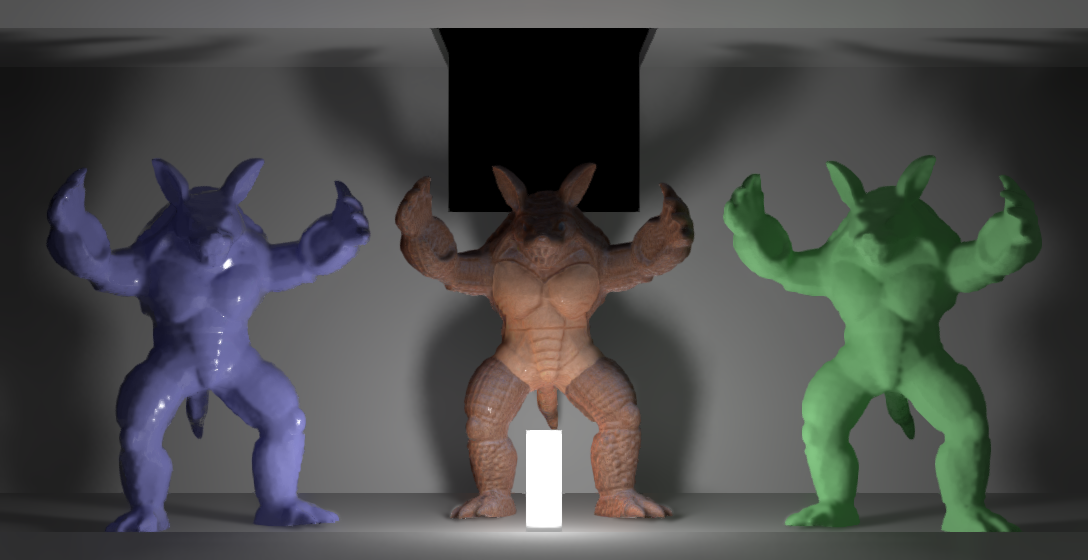}&
         \includegraphics[width=0.2999997\linewidth]{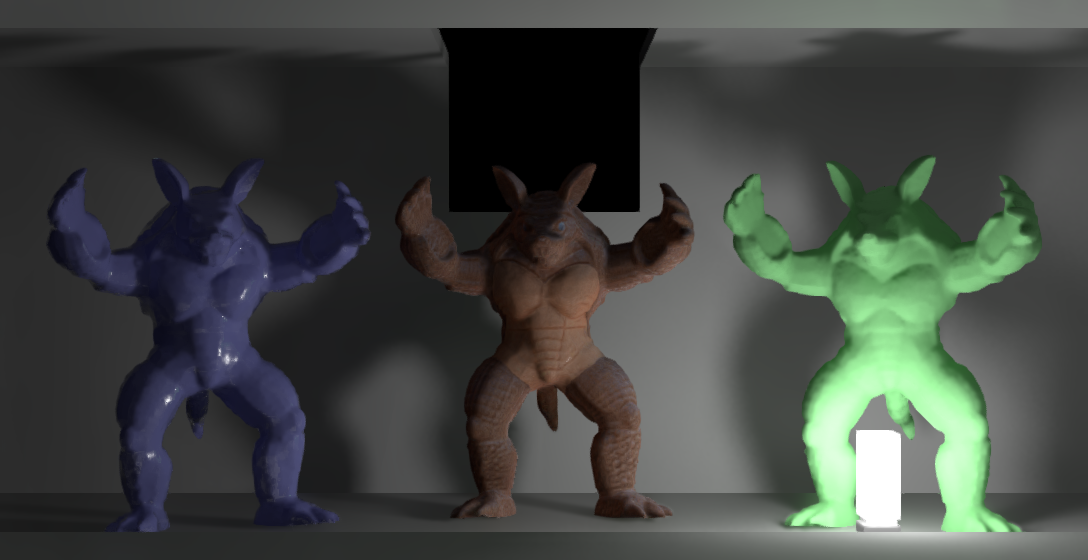}
         \\
         
         \includegraphics[width=0.2999997\linewidth]{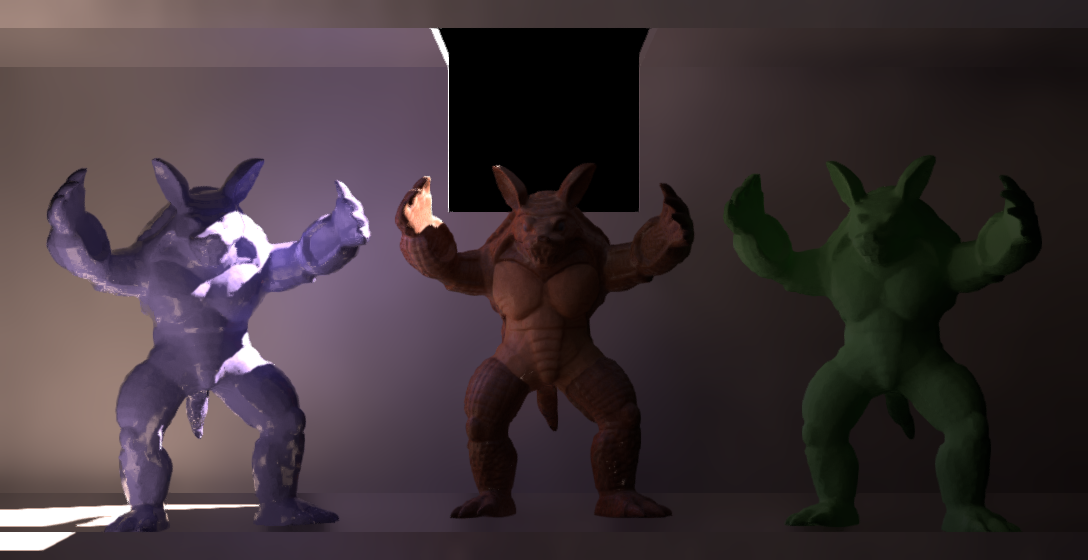}&
         \includegraphics[width=0.2999997\linewidth]{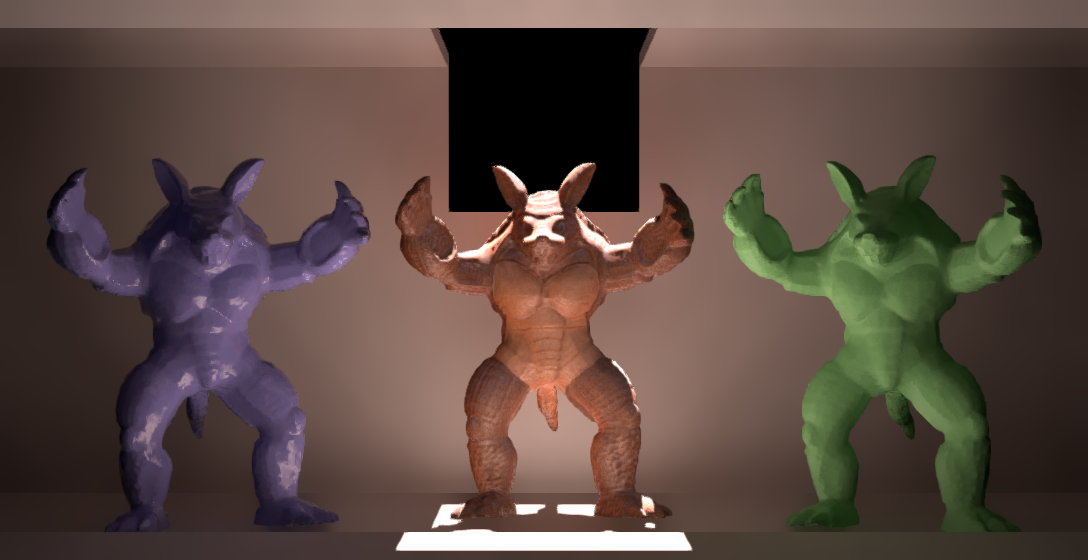}&
         \includegraphics[width=0.2999997\linewidth]{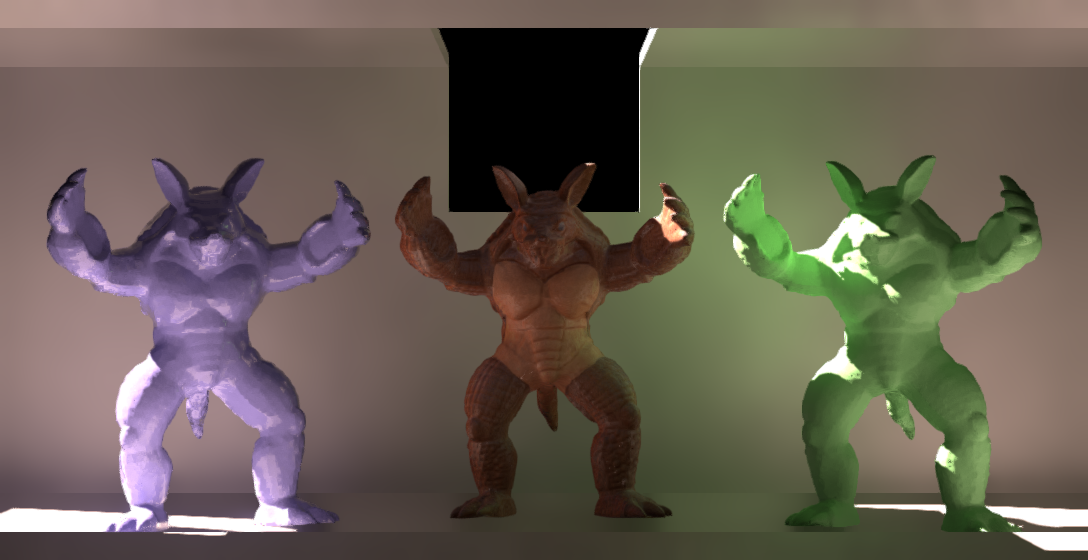}
    \end{tabular}
    \vspace{-0.2cm}
    \caption{First row: Relighting a single Armadillo with different lights.
    Second row: Relighting Armadillos with different area lights.
    Third row: Relighting Armadillos with different directional lights.
    All the lighting changes and relighting are completed dynamically in our application in real-time.}
    \label{fig:control-light}
\end{figure*}

\subsubsection{Various lighting conditions and material setting}

\begin{figure}
    \begin{tabular}{@{}c@{}c@{}c@{}c@{}}
    \multicolumn{2}{@{}c@{}}{
    \includegraphics[width=0.5\linewidth]{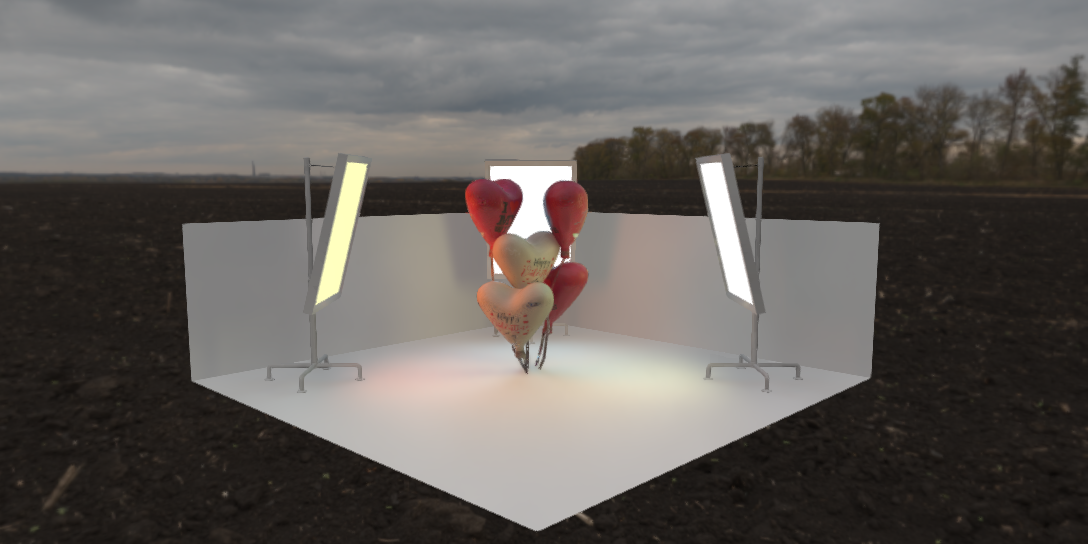}
    }&
    \includegraphics[width=0.25\linewidth]{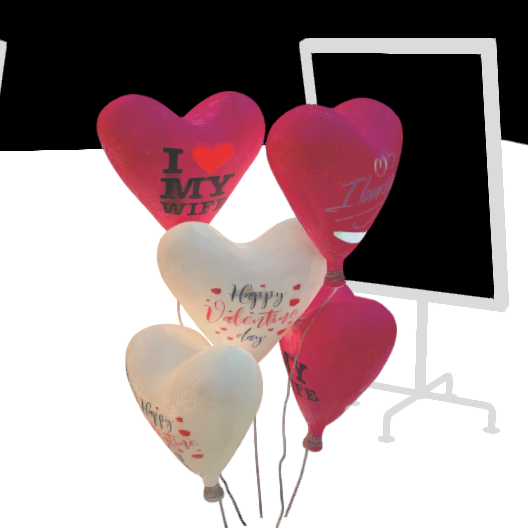}&
    \includegraphics[width=0.25\linewidth]{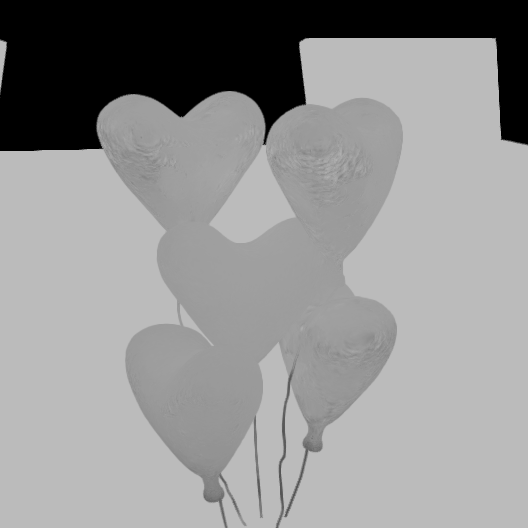}
    \\
    \multicolumn{2}{@{}c@{}}{Lighting Setting} &
    Albedo &
    Roughness
    \\
    \includegraphics[width=0.25\linewidth]{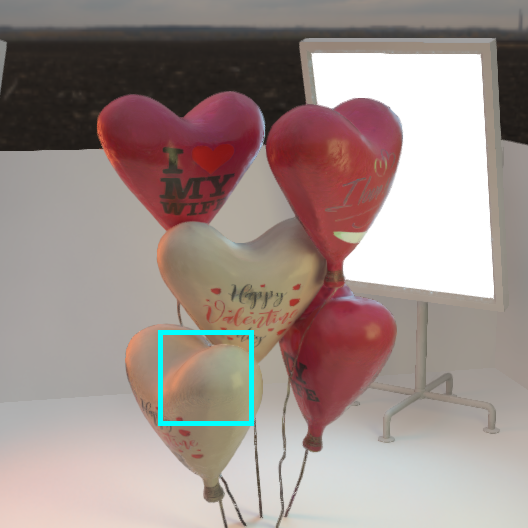}&
    \includegraphics[width=0.25\linewidth]{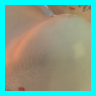}&
    \includegraphics[width=0.25\linewidth]{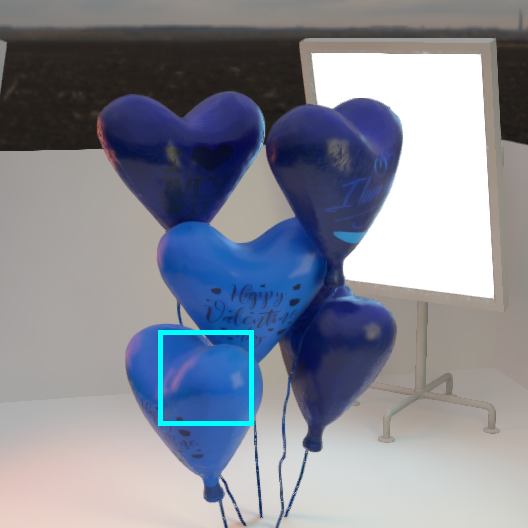}&
    \includegraphics[width=0.25\linewidth]{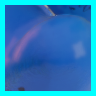}
    \\
    \multicolumn{2}{@{}c@{}}{(a) Original} &
    \multicolumn{2}{@{}c@{}}{(b) Bluish}
    \\
    \includegraphics[width=0.25\linewidth]{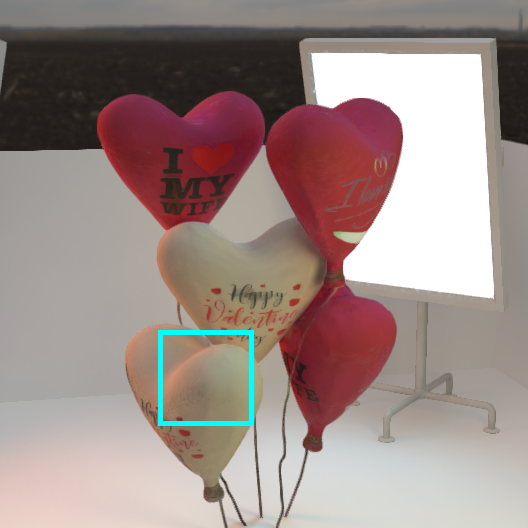}&
    \includegraphics[width=0.25\linewidth]{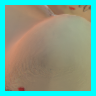}&
    \includegraphics[width=0.25\linewidth]{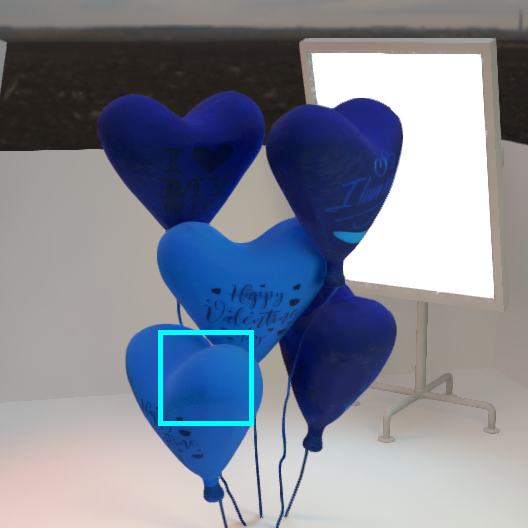}&
    \includegraphics[width=0.25\linewidth]{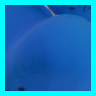}
    \\
    \multicolumn{2}{@{}c@{}}{(c) Rough} &
    \multicolumn{2}{@{}c@{}}{(d) Rough bluish}
    \\
    \includegraphics[width=0.25\linewidth]{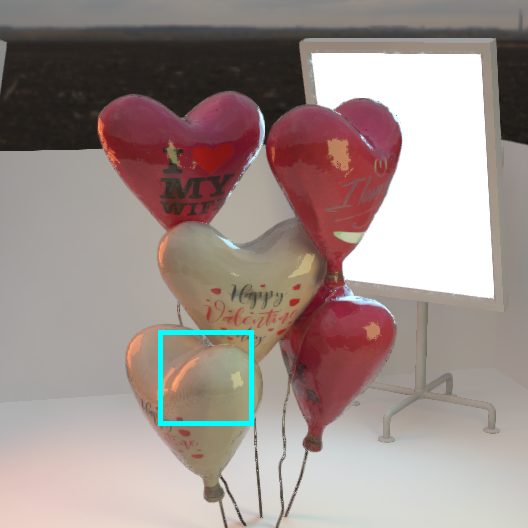}&
    \includegraphics[width=0.25\linewidth]{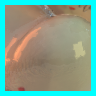}&
    \includegraphics[width=0.25\linewidth]{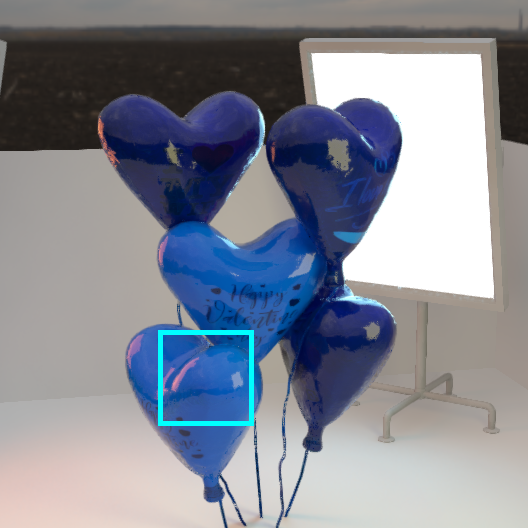}&
    \includegraphics[width=0.25\linewidth]{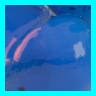}
    \\
    \multicolumn{2}{@{}c@{}}{(e) Smooth} &
    \multicolumn{2}{@{}c@{}}{(f) Smooth bluish} 
    \end{tabular}
    \caption{In-application material variation. We visualize the original material parameters and show the modified versions under the same lighting conditions.}
    \label{fig:control-material}
\end{figure}

We demonstrate our approach under various lighting conditions with many types of light sources (environment, directional, area), with different materials, and across multiple scenes, highlighting the real-time GI capabilities of our pipeline. 

For various lighting conditions, we present results for each light type individually while dynamically adjusting lighting parameters in \autoref{fig:control-light}. Our approach effectively renders multiple dynamic lights. Area lights produce soft shadows. Strong directional lights illuminate different Armadillos, scattering diffuse indirect lighting on the walls and other Armadillos.

\begin{figure}
    \centering
    \includegraphics[width=0.9\linewidth]{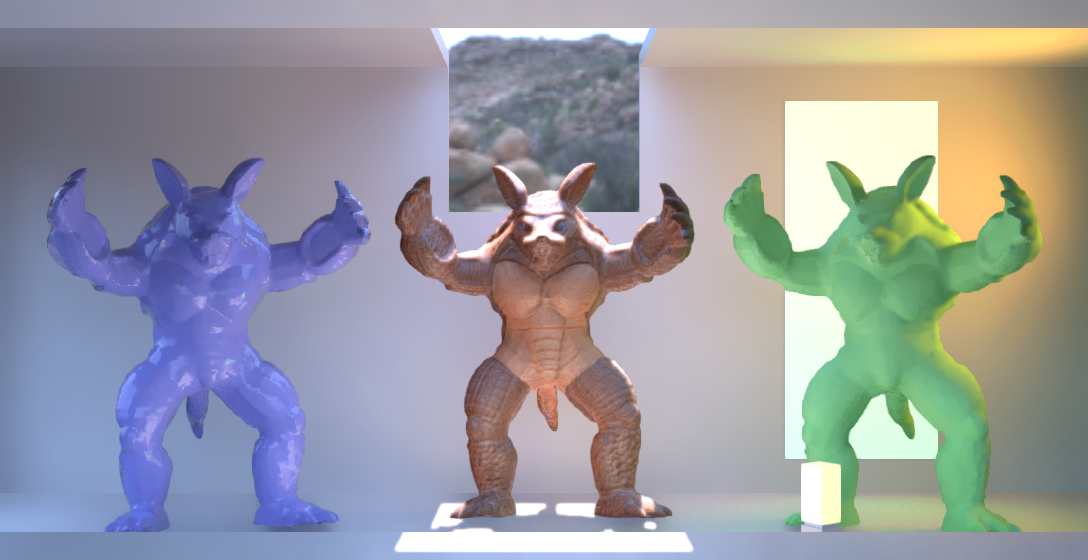}
    \caption{Three 3D Gaussian Armadillos with roughness variation. The blue glossy object (left) is lit by skylight through a wall opening, while the green and rough one (right) is illuminated by multiple area lights. Our RTGI can handle diffuse and glossy materials under various light settings.}
    \label{fig:materials-and-lights}
\end{figure}

For materials, we show the illumination produced by alternating material properties of 3D Gaussians dynamically in \autoref{fig:control-material}. Material parameters are alternated in our application in real time. Our approach successfully approximates the appearances of 3D Gaussian models with different colors and roughesses. 
\autoref{fig:materials-and-lights} shows the GI effect that combines different materials and all types of lighting. 

\begin{figure}
    \begin{tabular}{c@{}c@{}c@{}c}
         \includegraphics[width=0.5\linewidth]{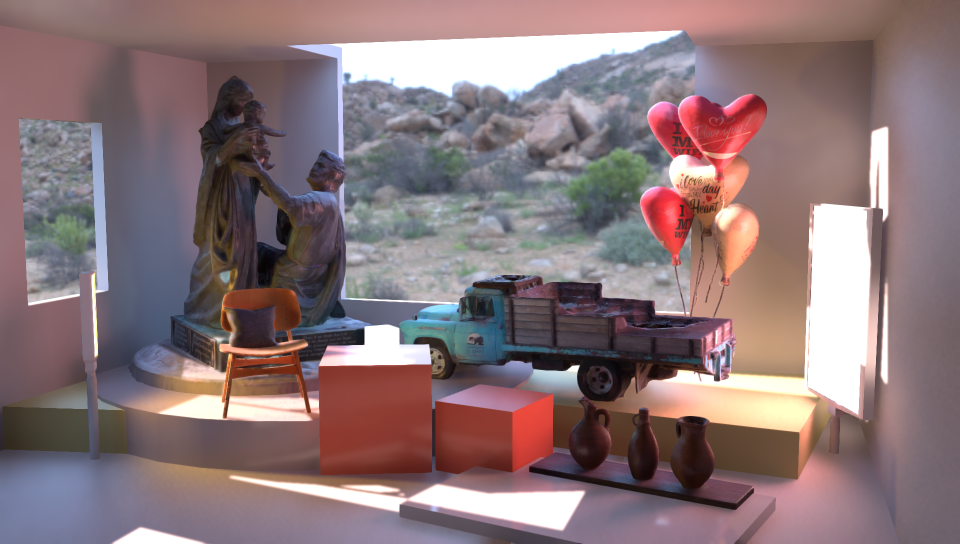}& 
         \includegraphics[width=0.5\linewidth]{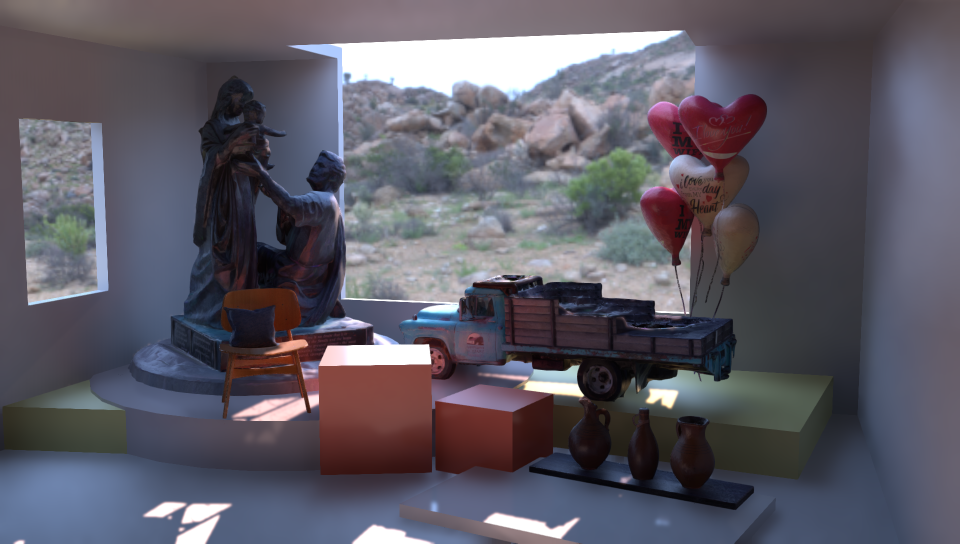}
         \\
         \includegraphics[width=0.5\linewidth]{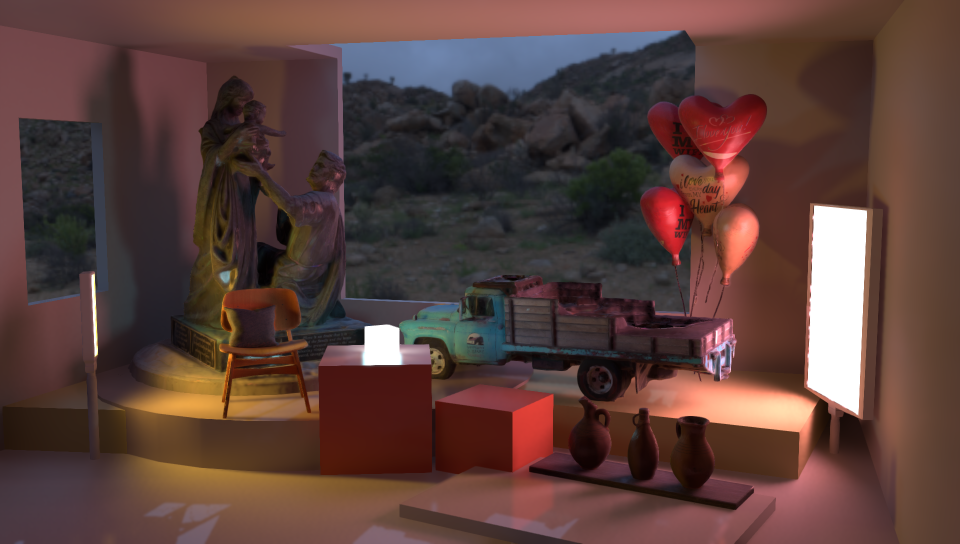}& 
         \includegraphics[width=0.5\linewidth]{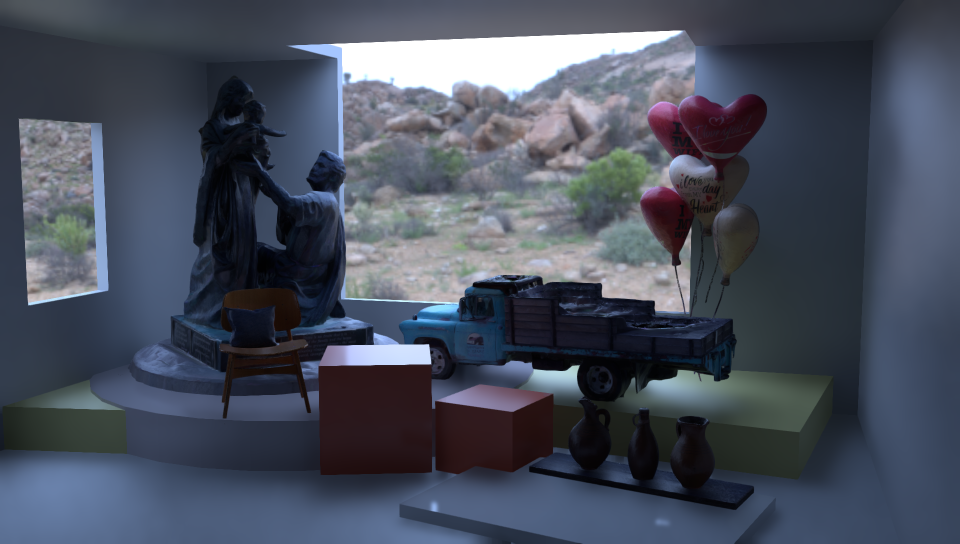}
         \\
          \includegraphics[width=0.5\linewidth]{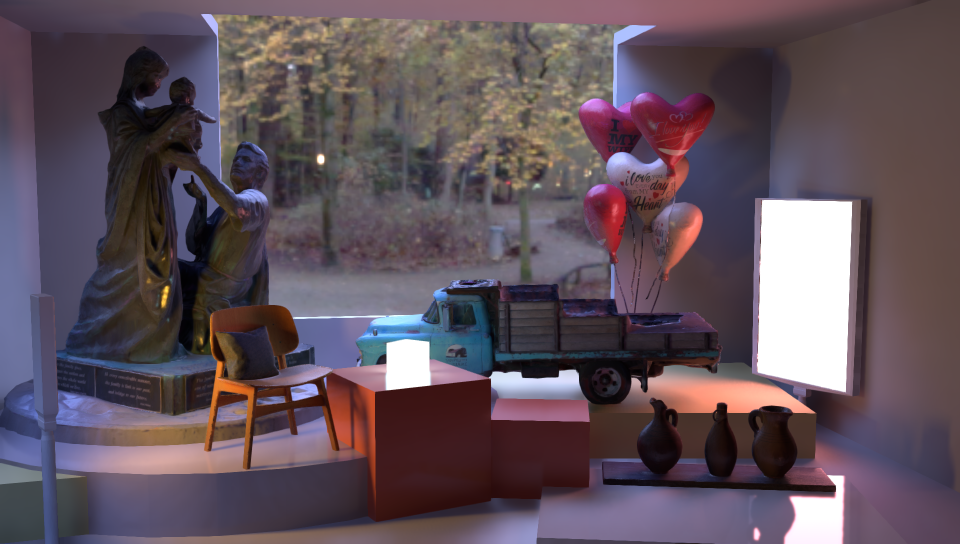}& 
         \includegraphics[width=0.5\linewidth]{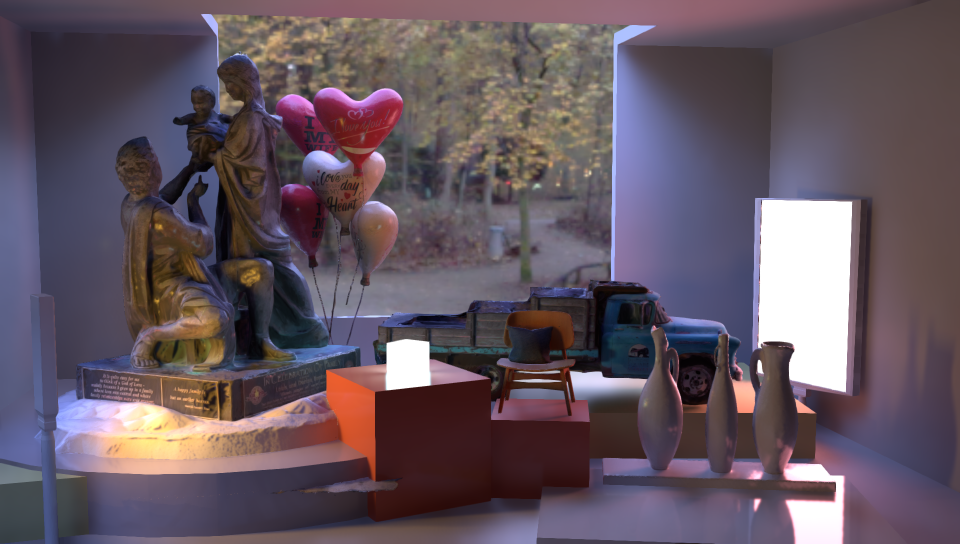}
         \\
         \includegraphics[width=0.5\linewidth]{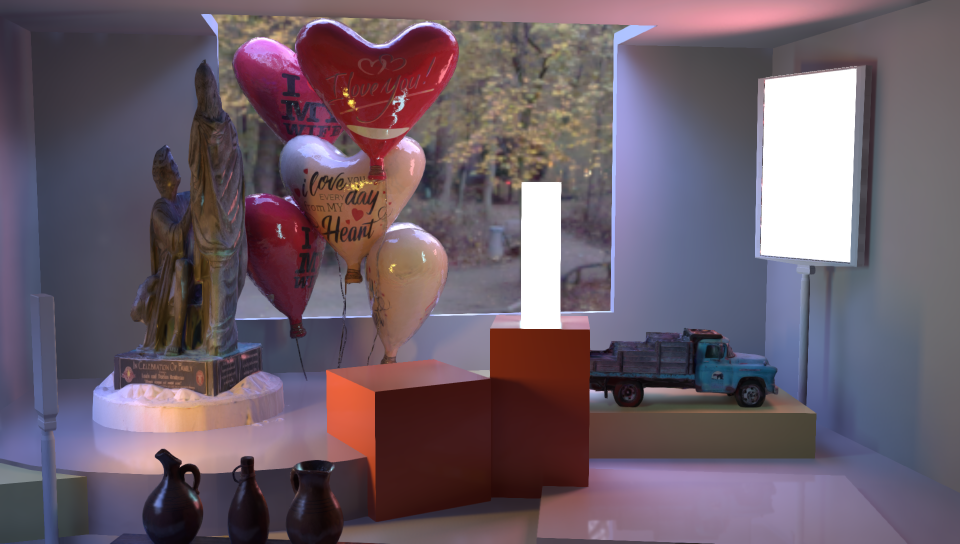}& 
         \includegraphics[width=0.5\linewidth]{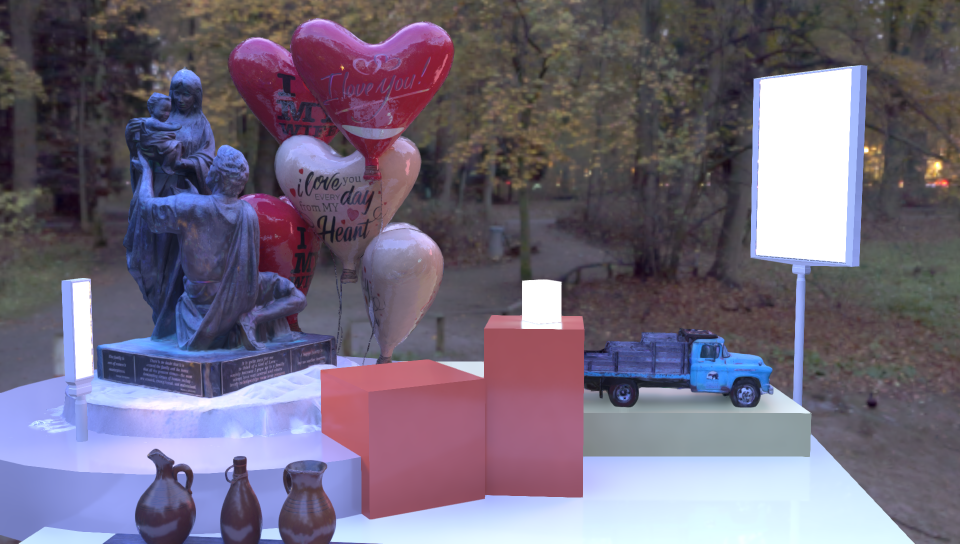}
    \end{tabular}
    \vspace{-0.3cm}
    \caption{Dynamic scene with complex light settings. Each is rendered with modified scene parameters in our interactive application. All the models and light sources can be transformed, scaled, inserted, or removed. In other words, the scene is fully interactive and editable in real time.}
    \label{fig:various-complex-lighting}
\end{figure}

To further validate our approach, we test it on the more complex teaser scene, showcasing relighting results under dynamically adjusted scene parameters in \autoref{fig:various-complex-lighting}. We freely modify the camera parameters, object transforms and scales, material properties, and lighting conditions. Our approach consistently delivers high-quality global illumination across all settings.

\begin{figure}
    \centering
    \includegraphics[width=\linewidth]{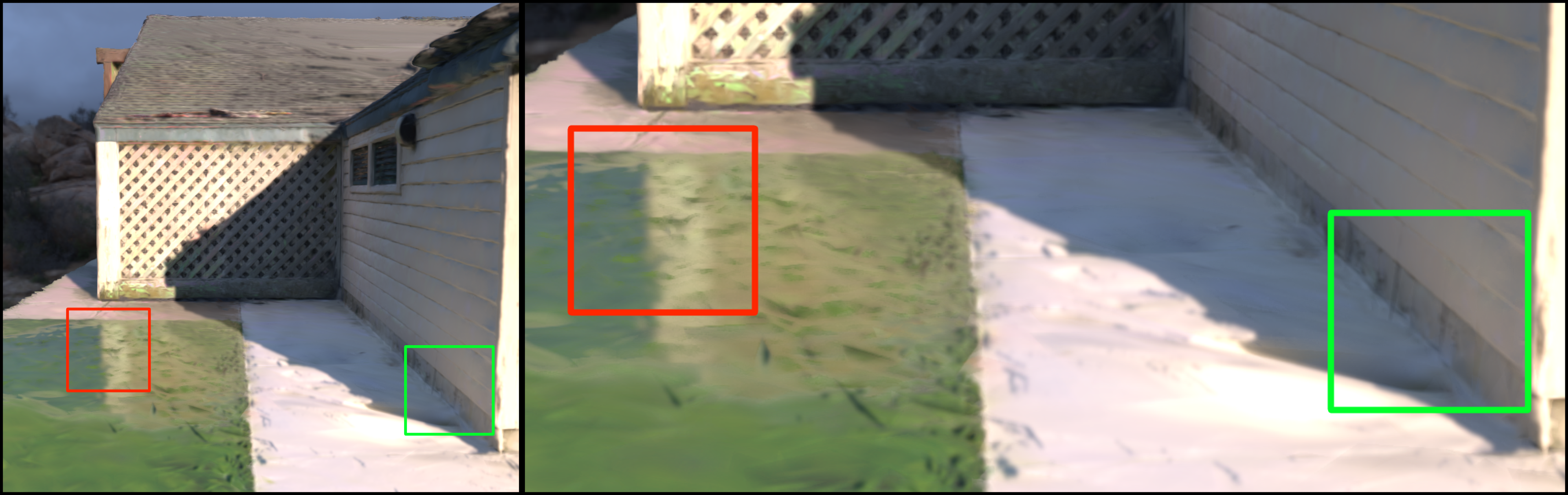}
    \caption{A 3D Gaussian barn scene lit by a directional light and a sky light. Light bouncing within the 3D Gaussian model itself is well captured (zoomed details on the right). The damp grass (red rectangle) reflects the barn walls, while the shadowed barn wall (green rectangle) receives scattered orange light from the sunlit walkways.}
    \label{fig:self-indirect}
\end{figure}

\begin{figure}
    \centering
    \begin{tabular}{@{}c@{\hspace{0.1cm}}c@{}}
        \includegraphics[width=0.48\linewidth]{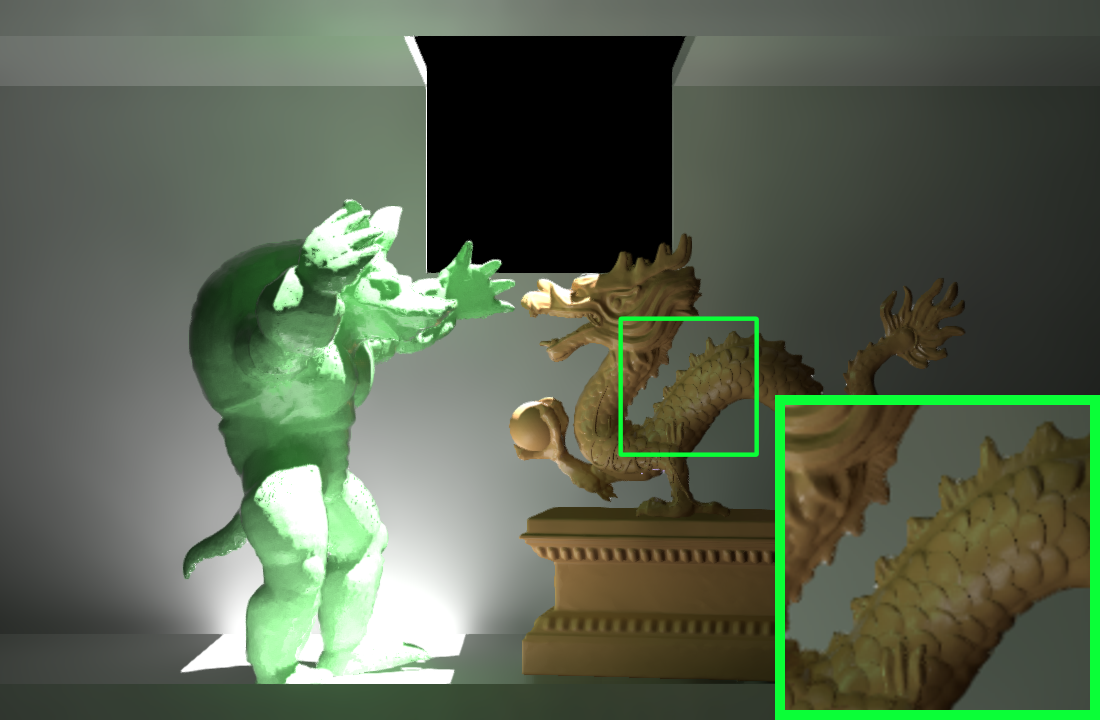} &
         \includegraphics[width=0.48\linewidth]{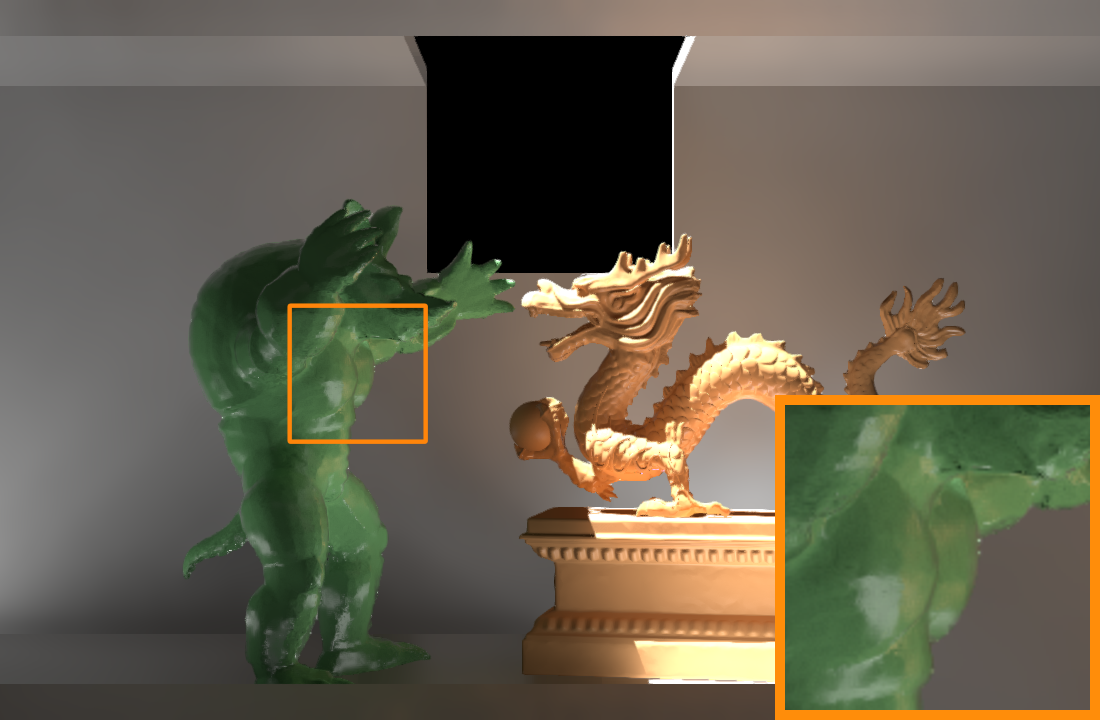} \\
         3D Gaussians $\rightarrow$ mesh &
         Mesh $\rightarrow$ 3D Gaussians
    \end{tabular}
    \vspace{-0.3cm}
    \caption{Mutual indirect light transport between 3D Gaussian model (Armadillo) and mesh (Dragon). In this scene, the directional light comes from the roof opening.} 
    \label{fig:mutual}
\end{figure}

We examine indirect illumination solely coming from light bouncing between 3D Gaussians, which can be hard to perceive without deliberate scene design, as shown in \autoref{fig:self-indirect}. Additionally, we demonstrate mutual indirect light transport between 3D Gaussians (Armadillo) and a mesh (Dragon) in \autoref{fig:mutual}. We observe that the Dragon model receives indirect lighting from the Armadillo, giving it a slight green tint, while the Armadillo, in turn, is indirectly lit by the Dragon, resulting in a subtle orange hue. These results highlight our approach’s ability to accurately capture indirect illumination transporting between 3D Gaussians and traditional mesh, seamlessly integrating them altogether in global illumination.

\subsubsection{Dynamic Scene}
By updating the scene lighting progressively, our approach handles dynamic scene changes without offline pre-computation. Each figure in \autoref{fig:various-complex-lighting} is generated in our application by dynamically adjusting scene parameters in real time, including object transformations, insertion/removal, as well as variations in material properties and lighting.
We provide a video in the supplements showing real-time navigating though the scene, and live editing of the object transformations, materials, and lights.

\begin{figure}
    \centering
    \includegraphics[width=\linewidth]{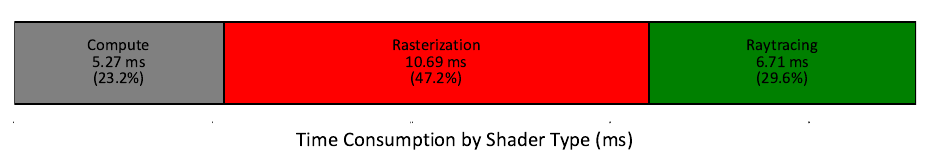}
    \vspace{-0.2cm}
    \includegraphics[width=\linewidth]{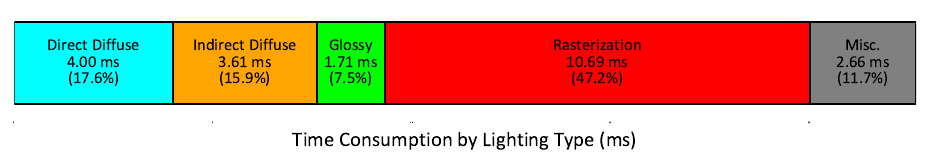}
    \caption{Time consumption of various shaders and lighting effects with single frame. Mesh rasterization (less than 0.1 ms) is excluded. Covariance matrix computation and depth sorting for 3D Gaussians are classified as compute or misc workloads.}
    \label{fig:perf}
    \vspace{0.5cm}
\end{figure}

\subsection{Performance}
We analyze our pipeline's performance using the complex teaser scene, containing 1.57 million 3D Gaussian primitives inside the view frustum, as shown in \autoref{fig:teaser}. Our approach achieves an average frame rate of 43 at 1920x1088 resolution with identical camera settings that are used to render the teaser. We plot the time consumption of different types of shaders and different lighting effects, as shown in \autoref{fig:perf}. For simple scenarios such as single object relighting, we can reach 200+ fps at a resolution of 1280x720, which is shown in the supplemented video.
Surprisingly, 3D Gaussian rasterization is the main performance bottleneck, scaling linearly with total fragment byte size. Ray-tracing cost grows with model size but in a sub-linear rate. Its cost remains minor compared to that of rasterization. Other computational costs stay largely constant. 

\begin{figure}
    \centering
    \begin{tabular}{c@{\hspace{0.1cm}}c}
        \includegraphics[width=0.4\linewidth]{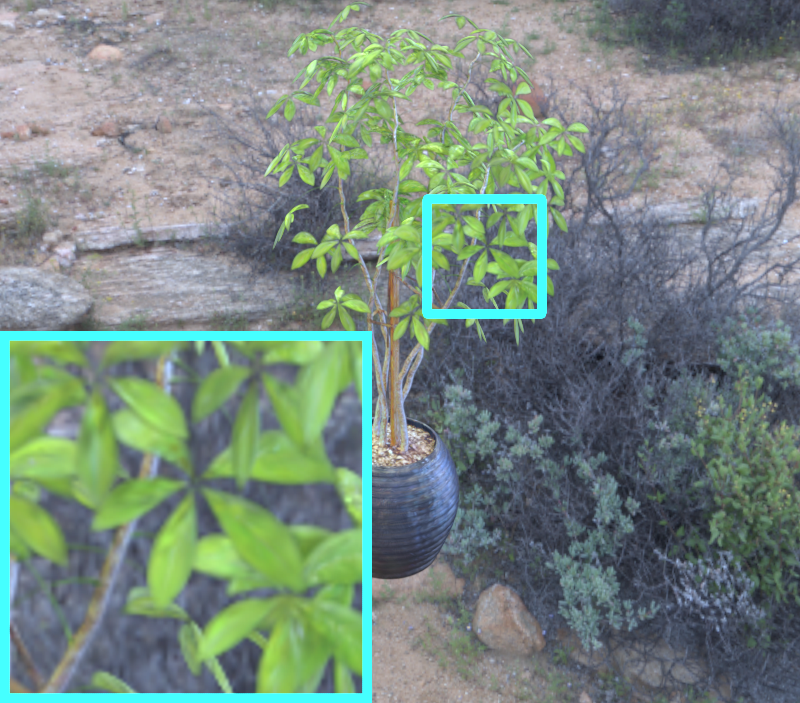} &
        \includegraphics[width=0.4\linewidth]{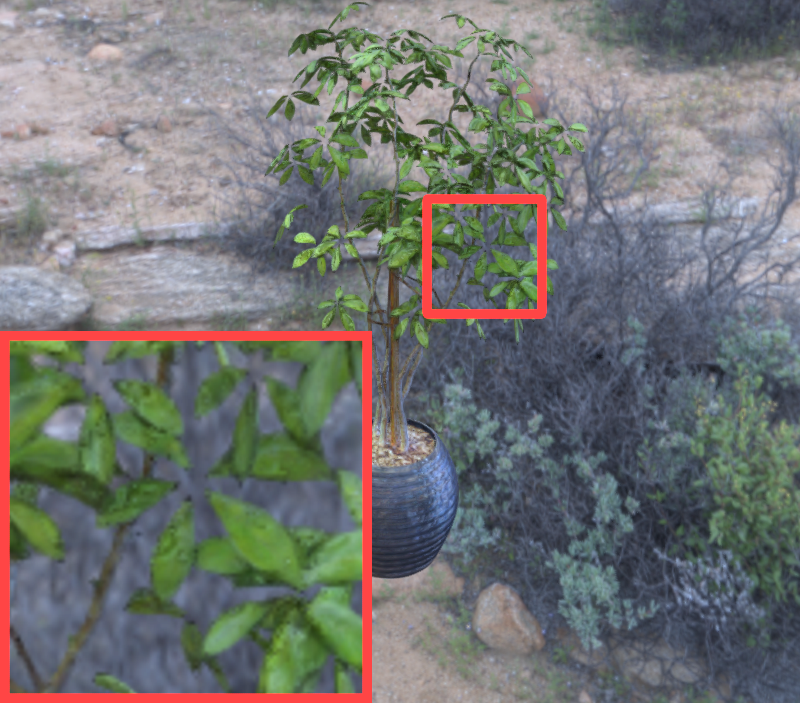}
    \end{tabular}
    \vspace{-0.3cm}
    \caption{A failure case on 3D Gaussians with complex and non-precise reconstructed geometries. The sparse screen probes fail to cover and shade all pixels due to misleading and low-quality reconstructed surfaces, ultimately darkening the output (right) compared to R3DG (left). } 
    \label{fig:fail-case-ficus}
\end{figure}


\section{Conclusion, Limitation, and Future Work}
In this paper, we define the global illumination problem for 3D Gaussians with a modified surface LTE, and present a RTGI approach for dynamic scenes and lighting settings, without pre-computations. 
By demonstrating the pipeline, we extend the capabilities of 3D Gaussian rendering, showcasing the feasibility of RTGI for 3D Gaussians. Additionally, we provide empirical insights into optimizing RTGI for 3D Gaussians and mitigating existing challenges.
Our study can serve as a valuable bridge between the rendering and vision communities, enhancing the visual fidelity of scenes derived from the real world. This work may lay the foundation for more realistic rendering techniques that leverage 3D Gaussian models and scenes, offering an alternative to traditional scene representations.

Our approach has several limitations, which we aim to address in future work.
First, it relies on well-formed 3D Gaussian models and strong assumptions. When these assumptions break down or partially fail, as seen in \autoref{fig:fail-case-ficus}, noticeable lighting artifacts occur. Therefore, improving geometry accuracy for 3D Gaussian models remains a research topic.
Second, as analyzed and shown in \autoref{fig:perf}, rasterization is a major performance bottleneck, with overdraw being the key limiting factor. Rendering 3DGS models heavily loads the GPU’s output merging and blending units, leading to drastic performance downgrades. Minimizing overdraw is essential for efficiency.
Third, our pipeline supports only a limited range of materials, assuming 3D Gaussian surfaces are well-formed. RTGI for transmissive materials and irregular surfaces remains unexplored. Notably, the 3D Gaussian reconstruction of translucent or transparent objects is a significant challenge in the field of computer vision.
Additionally, glossy reflectance from our pipeline may exhibit flickering. It's noticeable in the supplementary video, which could be mitigated with improved temporal anti-aliasing.

\bibliographystyle{ACM-Reference-Format}
\bibliography{references}

@InProceedings{gaoR3DG2023,
author="Gao, Jian
and Gu, Chun
and Lin, Youtian
and Li, Zhihao
and Zhu, Hao
and Cao, Xun
and Zhang, Li
and Yao, Yao",
editor="Leonardis, Ale{\v{s}}
and Ricci, Elisa
and Roth, Stefan
and Russakovsky, Olga
and Sattler, Torsten
and Varol, G{\"u}l",
title="Relightable 3D Gaussians: Realistic Point Cloud Relighting with BRDF Decomposition and Ray Tracing",
booktitle="Computer Vision -- ECCV 2024",
year="2025",
publisher="Springer Nature Switzerland",
address="Cham",
pages="73--89",
isbn="978-3-031-72995-9"
}

@article{nvidia3DGSRT2024,
author = {Moenne-Loccoz, Nicolas and Mirzaei, Ashkan and Perel, Or and de Lutio, Riccardo and Martinez Esturo, Janick and State, Gavriel and Fidler, Sanja and Sharp, Nicholas and Gojcic, Zan},
title = {3D Gaussian Ray Tracing: Fast Tracing of Particle Scenes},
year = {2024},
issue_date = {December 2024},
publisher = {Association for Computing Machinery},
address = {New York, NY, USA},
volume = {43},
number = {6},
issn = {0730-0301},
url = {https://doi.org/10.1145/3687934},
doi = {10.1145/3687934},
journal = {ACM Trans. Graph.},
month = nov,
articleno = {232},
numpages = {19}
}

@article{boisse2023GI10,
  title={GI-1.0: A Fast and Scalable Two-level Radiance Caching Scheme for Real-time Global Illumination},
  author={Boiss{\'e}, Guillaume and Meunier, Sylvain and de Dinechin, Heloise and Bartels, Pieterjan and Veselov, Alexander and Eto, Kenta and Harada, Takahiro},
  journal={arXiv preprint arXiv:2310.19855},
  year={2023}
}

@article{majercik2019DDGI,
  title={Dynamic diffuse global illumination with ray-traced irradiance fields},
  author={Majercik, Zander and Guertin, Jean-Philippe and Nowrouzezahrai, Derek and McGuire, Morgan},
  journal={Journal of Computer Graphics Techniques},
  volume={8},
  number={2},
  year={2019}
}

@inproceedings{ouyang2021ReSTIRGI,
  title={ReSTIR GI: Path resampling for real-time path tracing},
  author={Ouyang, Yaobin and Liu, Shiqiu and Kettunen, Markus and Pharr, Matt and Pantaleoni, Jacopo},
  booktitle={Computer Graphics Forum},
  volume={40},
  number={8},
  pages={17--29},
  year={2021},
  organization={Wiley Online Library}
}

@article{daqi2022GRIS,
author = {Lin, Daqi and Kettunen, Markus and Bitterli, Benedikt and Pantaleoni, Jacopo and Yuksel, Cem and Wyman, Chris},
title = {Generalized resampled importance sampling: foundations of ReSTIR},
year = {2022},
issue_date = {July 2022},
publisher = {Association for Computing Machinery},
address = {New York, NY, USA},
volume = {41},
number = {4},
issn = {0730-0301},
url = {https://doi.org/10.1145/3528223.3530158},
doi = {10.1145/3528223.3530158},
journal = {ACM Trans. Graph.},
month = jul,
articleno = {75},
numpages = {23}
}

@inproceedings{wright2022lumen,
  title={Lumen: Real-time global illumination in unreal engine 5},
  author={Wright, Daniel and Narkowicz, Krzysztof and Kelly, Patrick},
  booktitle={ACM SIGGRAPH},
  year={2022}
}

@article{bitterli2020ReSTIRDI,
author = {Bitterli, Benedikt and Wyman, Chris and Pharr, Matt and Shirley, Peter and Lefohn, Aaron and Jarosz, Wojciech},
title = {Spatiotemporal reservoir resampling for real-time ray tracing with dynamic direct lighting},
year = {2020},
issue_date = {August 2020},
publisher = {Association for Computing Machinery},
address = {New York, NY, USA},
volume = {39},
number = {4},
issn = {0730-0301},
url = {https://doi.org/10.1145/3386569.3392481},
doi = {10.1145/3386569.3392481},
journal = {ACM Trans. Graph.},
month = aug,
articleno = {148},
numpages = {17}
}

@inproceedings{crassin2011interactive,
	address = {New York, NY, USA},
	series = {{SIGGRAPH} '11},
	title = {Interactive indirect illumination using voxel-based cone tracing: an insight},
	isbn = {978-1-4503-0974-5},
	shorttitle = {Interactive indirect illumination using voxel-based cone tracing},
	url = {https://dl.acm.org/doi/10.1145/2037826.2037853},
	doi = {10.1145/2037826.2037853},
	urldate = {2025-01-08},
	booktitle = {{ACM} {SIGGRAPH} 2011 {Talks}},
	publisher = {Association for Computing Machinery},
	author = {Crassin, Cyril and Neyret, Fabrice and Sainz, Miguel and Green, Simon and Eisemann, Elmar},
	month = aug,
	year = {2011},
	pages = {1},
}

@article{mildenhall2020NeRF,
author = {Mildenhall, Ben and Srinivasan, Pratul P. and Tancik, Matthew and Barron, Jonathan T. and Ramamoorthi, Ravi and Ng, Ren},
title = {NeRF: representing scenes as neural radiance fields for view synthesis},
year = {2021},
issue_date = {January 2022},
publisher = {Association for Computing Machinery},
address = {New York, NY, USA},
volume = {65},
number = {1},
issn = {0001-0782},
url = {https://doi.org/10.1145/3503250},
doi = {10.1145/3503250},
journal = {Commun. ACM},
month = dec,
pages = {99–106},
numpages = {8}
}

@inproceedings{bi2020NeRFRelightable,
author = {Bi, Sai and Xu, Zexiang and Sunkavalli, Kalyan and Ha\v{s}an, Milo\v{s} and Hold-Geoffroy, Yannick and Kriegman, David and Ramamoorthi, Ravi},
title = {Deep Reflectance Volumes: Relightable Reconstructions from Multi-view Photometric Images},
year = {2020},
isbn = {978-3-030-58579-2},
publisher = {Springer-Verlag},
address = {Berlin, Heidelberg},
url = {https://doi.org/10.1007/978-3-030-58580-8_18},
doi = {10.1007/978-3-030-58580-8_18},
booktitle = {Computer Vision – ECCV 2020: 16th European Conference, Glasgow, UK, August 23–28, 2020, Proceedings, Part III},
pages = {294–311},
numpages = {18},
location = {Glasgow, United Kingdom}
}

@inproceedings{jin2023Tensoir,
  title={Tensoir: Tensorial inverse rendering},
  author={Jin, Haian and Liu, Isabella and Xu, Peijia and Zhang, Xiaoshuai and Han, Songfang and Bi, Sai and Zhou, Xiaowei and Xu, Zexiang and Su, Hao},
  booktitle={Proceedings of the IEEE/CVF Conference on Computer Vision and Pattern Recognition},
  pages={165--174},
  year={2023}
}

@inproceedings{boss2021nerd,
  title={Nerd: Neural reflectance decomposition from image collections},
  author={Boss, Mark and Braun, Raphael and Jampani, Varun and Barron, Jonathan T and Liu, Ce and Lensch, Hendrik},
  booktitle={Proceedings of the IEEE/CVF International Conference on Computer Vision},
  pages={12684--12694},
  year={2021}
}

@inproceedings{barron2021mip,
  title={Mip-nerf: A multiscale representation for anti-aliasing neural radiance fields},
  author={Barron, Jonathan T and Mildenhall, Ben and Tancik, Matthew and Hedman, Peter and Martin-Brualla, Ricardo and Srinivasan, Pratul P},
  booktitle={Proceedings of the IEEE/CVF international conference on computer vision},
  pages={5855--5864},
  year={2021}
}

@article{muller2022instant,
  title={Instant neural graphics primitives with a multiresolution hash encoding},
  author={M{\"u}ller, Thomas and Evans, Alex and Schied, Christoph and Keller, Alexander},
  journal={ACM transactions on graphics (TOG)},
  volume={41},
  number={4},
  pages={1--15},
  year={2022},
  publisher={ACM New York, NY, USA}
}

@article{kerbl20233dgs,
  title={3d gaussian splatting for real-time radiance field rendering.},
  author={Kerbl, Bernhard and Kopanas, Georgios and Leimk{\"u}hler, Thomas and Drettakis, George},
  journal={ACM Trans. Graph.},
  volume={42},
  number={4},
  pages={139--1},
  year={2023}
}

@article{feng2024flashgs,
  title={Flashgs: Efficient 3d gaussian splatting for large-scale and high-resolution rendering},
  author={Feng, Guofeng and Chen, Siyan and Fu, Rong and Liao, Zimu and Wang, Yi and Liu, Tao and Pei, Zhilin and Li, Hengjie and Zhang, Xingcheng and Dai, Bo},
  journal={arXiv preprint arXiv:2408.07967},
  year={2024}
}

@article{wang2024pygs,
  title={PyGS: Large-scale Scene Representation with Pyramidal 3D Gaussian Splatting},
  author={Wang, Zipeng and Xu, Dan},
  journal={arXiv preprint arXiv:2405.16829},
  year={2024}
}

@inproceedings{liu2025citygaussian,
  title={Citygaussian: Real-time high-quality large-scale scene rendering with gaussians},
  author={Liu, Yang and Luo, Chuanchen and Fan, Lue and Wang, Naiyan and Peng, Junran and Zhang, Zhaoxiang},
  booktitle={European Conference on Computer Vision},
  pages={265--282},
  year={2025},
  organization={Springer}
}

@article{zwicker2002ewa,
  title={EWA splatting},
  author={Zwicker, Matthias and Pfister, Hanspeter and Van Baar, Jeroen and Gross, Markus},
  journal={IEEE Transactions on Visualization and Computer Graphics},
  volume={8},
  number={3},
  pages={223--238},
  year={2002},
  publisher={IEEE}
}

@inproceedings{yu2024mip,
  title={Mip-splatting: Alias-free 3d gaussian splatting},
  author={Yu, Zehao and Chen, Anpei and Huang, Binbin and Sattler, Torsten and Geiger, Andreas},
  booktitle={Proceedings of the IEEE/CVF Conference on Computer Vision and Pattern Recognition},
  pages={19447--19456},
  year={2024}
}

@inproceedings{liang2024gsir,
  title={Gs-ir: 3d gaussian splatting for inverse rendering},
  author={Liang, Zhihao and Zhang, Qi and Feng, Ying and Shan, Ying and Jia, Kui},
  booktitle={Proceedings of the IEEE/CVF Conference on Computer Vision and Pattern Recognition},
  pages={21644--21653},
  year={2024}
}

@inproceedings{liu2025mirrorgaussian,
  title={Mirrorgaussian: Reflecting 3d gaussians for reconstructing mirror reflections},
  author={Liu, Jiayue and Tang, Xiao and Cheng, Freeman and Yang, Roy and Li, Zhihao and Liu, Jianzhuang and Huang, Yi and Lin, Jiaqi and Liu, Shiyong and Wu, Xiaofei and others},
  booktitle={European Conference on Computer Vision},
  pages={377--393},
  year={2025},
  organization={Springer}
}

@inproceedings{bi2024gs3,
  title={Gs3: Efficient relighting with triple gaussian splatting},
  author={Bi, Zoubin and Zeng, Yixin and Zeng, Chong and Pei, Fan and Feng, Xiang and Zhou, Kun and Wu, Hongzhi},
  booktitle={SIGGRAPH Asia 2024 Conference Papers},
  pages={1--12},
  year={2024}
}

@article{fan2024rng,
  title={RNG: Relightable Neural Gaussians},
  author={Fan, Jiahui and Luan, Fujun and Yang, Jian and Ha{\v{s}}an, Milo{\v{s}} and Wang, Beibei},
  journal={arXiv preprint arXiv:2409.19702},
  year={2024}
}

@inproceedings{zeng2023nrhints,
  title={Relighting neural radiance fields with shadow and highlight hints},
  author={Zeng, Chong and Chen, Guojun and Dong, Yue and Peers, Pieter and Wu, Hongzhi and Tong, Xin},
  booktitle={ACM SIGGRAPH 2023 Conference Proceedings},
  pages={1--11},
  year={2023}
}

@inproceedings{jiang2024gaussianshader,
  title={Gaussianshader: 3d gaussian splatting with shading functions for reflective surfaces},
  author={Jiang, Yingwenqi and Tu, Jiadong and Liu, Yuan and Gao, Xifeng and Long, Xiaoxiao and Wang, Wenping and Ma, Yuexin},
  booktitle={Proceedings of the IEEE/CVF Conference on Computer Vision and Pattern Recognition},
  pages={5322--5332},
  year={2024}
}

@article{wu2024deferredgs,
  title={DeferredGS: Decoupled and Editable Gaussian Splatting with Deferred Shading},
  author={Wu, Tong and Sun, Jia-Mu and Lai, Yu-Kun and Ma, Yuewen and Kobbelt, Leif and Gao, Lin},
  journal={arXiv preprint arXiv:2404.09412},
  year={2024}
}

@inproceedings{wang2024adr,
  title={AdR-Gaussian: Accelerating Gaussian Splatting with Adaptive Radius},
  author={Wang, Xinzhe and Yi, Ran and Ma, Lizhuang},
  booktitle={SIGGRAPH Asia 2024 Conference Papers},
  pages={1--10},
  year={2024}
}

@article{efraimidis_weighted_2006,
	title = {Weighted random sampling with a reservoir},
	volume = {97},
	issn = {0020-0190},
	url = {https://www.sciencedirect.com/science/article/pii/S002001900500298X},
	doi = {https://doi.org/10.1016/j.ipl.2005.11.003},
	number = {5},
	journal = {Information Processing Letters},
	author = {Efraimidis, Pavlos S. and Spirakis, Paul G.},
	year = {2006},
	pages = {181--185},
}

@article{zhou2024unified,
      title={Unified Gaussian Primitives for Scene Representation and Rendering}, 
      author={Yang Zhou and Songyin Wu and Ling-Qi Yan},
      year={2024},
      eprint={2406.09733},
      archivePrefix={arXiv},
}

@article{condor2024dontsplat,
author = {Condor, Jorge and Speierer, Sebastien and Bode, Lukas and Bozic, Aljaz and Green, Simon and Didyk, Piotr and Jarabo, Adrian},
title = {Don't Splat your Gaussians: Volumetric Ray-Traced Primitives for Modeling and Rendering Scattering and Emissive Media},
year = {2025},
issue_date = {February 2025},
publisher = {Association for Computing Machinery},
address = {New York, NY, USA},
volume = {44},
number = {1},
issn = {0730-0301},
url = {https://doi.org/10.1145/3711853},
doi = {10.1145/3711853},
journal = {ACM Trans. Graph.},
month = feb,
articleno = {10},
numpages = {17}
}

@inproceedings{kaplanyan2010lpv,
  title={Cascaded light propagation volumes for real-time indirect illumination},
  author={Kaplanyan, Anton and Dachsbacher, Carsten},
  booktitle={Proceedings of the 2010 ACM SIGGRAPH symposium on Interactive 3D Graphics and Games},
  pages={99--107},
  year={2010}
}

@inproceedings{ye20243dgsreflection,
  title={3D Gaussian Splatting with Deferred Reflection},
  author={Ye, Keyang and Hou, Qiming and Zhou, Kun},
  booktitle={ACM SIGGRAPH 2024 Conference Papers},
  pages={1--10},
  year={2024}
}

@article{chen2024gi,
  title={GI-GS: Global Illumination Decomposition on Gaussian Splatting for Inverse Rendering},
  author={Chen, Hongze and Lin, Zehong and Zhang, Jun},
  journal={arXiv preprint arXiv:2410.02619},
  year={2024}
}

@online{lumaai,
    author       = {Luma AI},
    title        = {Luma AI - Interactive Scenes},
    year         = {2024},
    url          = {https://lumalabs.ai/interactive-scenes},
    note         = {Accessed: 2025-1-18},
}

@inproceedings{kajiya1986rendering,
  title={The rendering equation},
  author={Kajiya, James T},
  booktitle={Proceedings of the 13th annual conference on Computer graphics and interactive techniques},
  pages={143--150},
  year={1986}
}

@article{lambru2021RTGIanalysis,

  author={Lambru, Cristian and Morar, Anca and Moldoveanu, Florica and Asavei, Victor and Moldoveanu, Alin},

  journal={IEEE Access}, 

  title={Comparative Analysis of Real-Time Global Illumination Techniques in Current Game Engines}, 

  year={2021},

  volume={9},

  number={},

  pages={125158-125183},

  doi={10.1109/ACCESS.2021.3109663}}

@Inbook{boksansky2021manylightsGridRes,
author="Boksansky, Jakub
and Jukarainen, Paula
and Wyman, Chris",
editor="Marrs, Adam
and Shirley, Peter
and Wald, Ingo",
title="Rendering Many Lights with Grid-Based Reservoirs",
bookTitle="Ray Tracing Gems II: Next Generation Real-Time Rendering with DXR, Vulkan, and OptiX",
year="2021",
publisher="Apress",
address="Berkeley, CA",
pages="351--365",
isbn="978-1-4842-7185-8",
doi="10.1007/978-1-4842-7185-8_23",
url="https://doi.org/10.1007/978-1-4842-7185-8_23"
}

@article{karis2013real,
  title={Real shading in unreal engine 4},
  author={Karis, Brian and Games, Epic},
  journal={Proc. Physically Based Shading Theory Practice},
  volume={4},
  number={3},
  pages={1},
  year={2013}
}

@inproceedings{guo2024prtgs,
author = {Guo, Yijia and Bai, Yuanxi and Hu, Liwen and Guo, Ziyi and Liu, Mianzhi and Cai, Yu and Huang, Tiejun and Ma, Lei},
title = {PRTGS: Precomputed Radiance Transfer of Gaussian Splats for Real-Time High-Quality Relighting},
year = {2024},
isbn = {9798400706868},
publisher = {Association for Computing Machinery},
address = {New York, NY, USA},
url = {https://doi.org/10.1145/3664647.3680893},
doi = {10.1145/3664647.3680893},
booktitle = {Proceedings of the 32nd ACM International Conference on Multimedia},
pages = {5112–5120},
numpages = {9},
location = {Melbourne VIC, Australia},
series = {MM '24}
}

@book{veach1998robust,
  title={Robust Monte Carlo methods for light transport simulation},
  author={Veach, Eric},
  year={1998},
  publisher={Stanford University}
}

@inproceedings{jensen1996pm,
  title={Global illumination using photon maps},
  author={Jensen, Henrik Wann},
  booktitle={Eurographics workshop on Rendering techniques},
  pages={21--30},
  year={1996},
  organization={Springer}
}

@incollection{hachisuka2008ppm,
  title={Progressive photon mapping},
  author={Hachisuka, Toshiya and Ogaki, Shinji and Jensen, Henrik Wann},
  booktitle={ACM SIGGRAPH Asia 2008 papers},
  pages={1--8},
  year={2008}
}

@article{kaplanyan2013appm,
  title={Adaptive progressive photon mapping},
  author={Kaplanyan, Anton S and Dachsbacher, Carsten},
  journal={ACM Transactions on Graphics (TOG)},
  volume={32},
  number={2},
  pages={1--13},
  year={2013},
  publisher={ACM New York, NY, USA}
}

@article{lin2020cppm,
  title={CPPM: chi-squared progressive photon mapping},
  author={Lin, Zehui and Li, Sheng and Zeng, Xinlu and Zhang, Congyi and Jia, Jinzhu and Wang, Guoping and Manocha, Dinesh},
  journal={ACM Transactions on Graphics (TOG)},
  volume={39},
  number={6},
  pages={1--12},
  year={2020},
  publisher={ACM New York, NY, USA}
}

@article{lin2023fvcm,
  title={Hypothesis Testing for Progressive Kernel Estimation and VCM Framework},
  author={Lin, Zehui and Hu, Chenxiao and Jia, Jinzhu and Li, Sheng},
  journal={IEEE Transactions on Visualization and Computer Graphics},
  volume={30},
  number={8},
  pages={4709--4723},
  year={2023},
  publisher={IEEE}
}

@inproceedings{popov2015probabilistic,
  title={Probabilistic connections for bidirectional path tracing},
  author={Popov, Stefan and Ramamoorthi, Ravi and Durand, Fredo and Drettakis, George},
  booktitle={Computer Graphics Forum},
  volume={34},
  number={4},
  pages={75--86},
  year={2015},
  organization={Wiley Online Library}
}

@article{su2022spcbpt,
  title={SPCBPT: Subspace-based probabilistic connections for bidirectional path tracing},
  author={Su, Fujia and Li, Sheng and Wang, Guoping},
  journal={ACM Transactions on Graphics (TOG)},
  volume={41},
  number={4},
  pages={1--14},
  year={2022},
  publisher={ACM New York, NY, USA}
}

@article{su2025proxy,
  title={Proxy Tracing: Unbiased Reciprocal Estimation for Optimized Sampling in BDPT},
  author={Su, Fujia and Li, Bingxuan and Yin, Qingyang and Zhang, Yanchen and Li, Sheng},
  journal={ACM Transactions on Graphics (TOG)},
  volume={43},
  number={4},
  pages={1--21},
  year={2024},
  publisher={ACM New York, NY, USA}
}

@inproceedings{dong2023neural,
  title={Neural parametric mixtures for path guiding},
  author={Dong, Honghao and Wang, Guoping and Li, Sheng},
  booktitle={ACM SIGGRAPH 2023 Conference Proceedings},
  pages={1--10},
  year={2023}
}

@article{huang2024online,
  title={Online Neural Path Guiding with Normalized Anisotropic Spherical Gaussians},
  author={Huang, Jiawei and Iizuka, Akito and Tanaka, Hajime and Komura, Taku and Kitamura, Yoshifumi},
  journal={ACM Transactions on Graphics},
  volume={43},
  number={3},
  pages={1--18},
  year={2024},
  publisher={ACM New York, NY}
}

@inproceedings{muller2017practical,
  title={Practical path guiding for efficient light-transport simulation},
  author={M{\"u}ller, Thomas and Gross, Markus and Nov{\'a}k, Jan},
  booktitle={Computer Graphics Forum},
  volume={36},
  number={4},
  pages={91--100},
  year={2017},
  organization={Wiley Online Library}
}

@article{ruppert2020robust,
  title={Robust fitting of parallax-aware mixtures for path guiding},
  author={Ruppert, Lukas and Herholz, Sebastian and Lensch, Hendrik PA},
  journal={ACM Transactions on Graphics (TOG)},
  volume={39},
  number={4},
  pages={147--1},
  year={2020},
  publisher={ACM New York, NY, USA}
}

@incollection{hachisuka2008multidimensional,
  title={Multidimensional adaptive sampling and reconstruction for ray tracing},
  author={Hachisuka, Toshiya and Jarosz, Wojciech and Weistroffer, Richard Peter and Dale, Kevin and Humphreys, Greg and Zwicker, Matthias and Jensen, Henrik Wann},
  booktitle={ACM SIGGRAPH 2008 papers},
  pages={1--10},
  year={2008}
}

@article{rousselle2012adaptive,
  title={Adaptive rendering with non-local means filtering},
  author={Rousselle, Fabrice and Knaus, Claude and Zwicker, Matthias},
  journal={ACM Transactions on Graphics (TOG)},
  volume={31},
  number={6},
  pages={1--11},
  year={2012},
  publisher={ACM New York, NY, USA}
}

@article{rousselle2011greedy,
  title={Adaptive sampling and reconstruction using greedy error minimization},
  author={Rousselle, Fabrice and Knaus, Claude and Zwicker, Matthias},
  journal={ACM Transactions on Graphics (TOG)},
  volume={30},
  number={6},
  pages={1--12},
  year={2011},
  publisher={ACM New York, NY, USA}
}

@article{salehi2022deep,
  title={Deep adaptive sampling and reconstruction using analytic distributions},
  author={Salehi, Farnood and Manzi, Marco and Roethlin, Gerhard and Weber, Romann and Schroers, Christopher and Papas, Marios},
  journal={ACM Transactions on Graphics (TOG)},
  volume={41},
  number={6},
  pages={1--16},
  year={2022},
  publisher={ACM New York, NY, USA}
}
\appendix

\noindent \textbf{Property:} $E(b) = 1 - T(r)$, where $b$ is the expectation for the shadow ray tracing result, and $T(r)$ is the ray translucency the reference algorithm evaluates to. 

\begin{proof}
For the reference algorithm modified from \cite{nvidia3DGSRT2024}, the ray translucency $T(r) =\prod_{g\in \mathrm{G}} (1 - A_{g, \mathbf{v}}(r))$ where $G$ stands for the set of 3D Gaussians whose proxy geometry the ray intersects with.
For our algorithm, the probability of accepting a hit upon intersecting the proxy geometry of a Gaussian is $A_{\mathbf{v},g}(r)$.

The discrete probability $E(b=0)$ for the entire shadow ray rejecting all hits, is $\prod_{g\in \mathrm{G}} (1 - A_{\mathbf{v}, g}(r)) = T(r)$, which is exactly the ray translucency the reference algorithm evaluates to. Thus, for the expectation of $b$, we have:
$\mathrm{E}(b) = 1 \times E(b=1) + 0\times E(b=0) = (1 - E(b=0)) = 1 - T(r)$.
\end{proof}

\begin{figure}[t]
    \centering
    \includegraphics[width=0.85\linewidth]{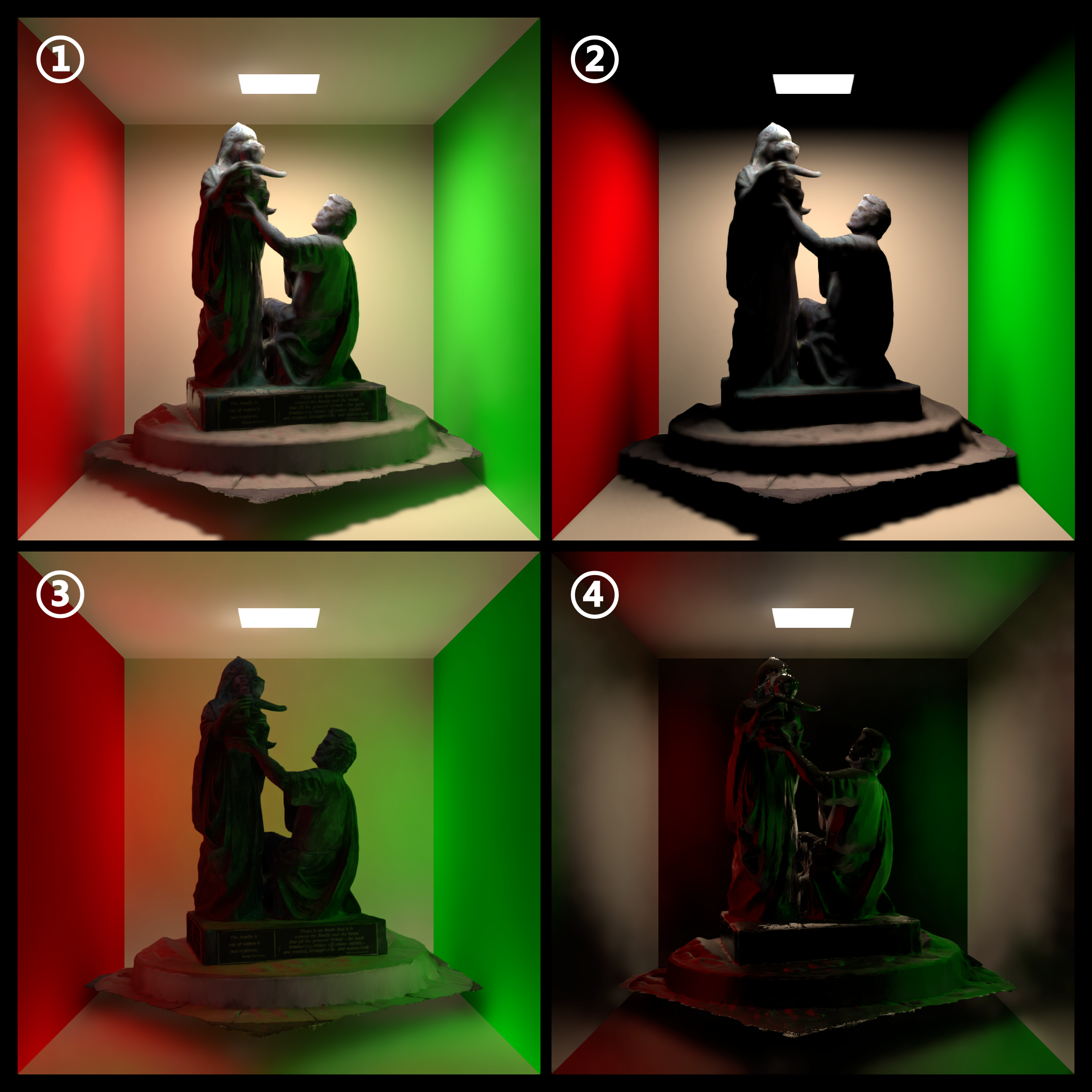}
    \caption{"Family" model in Cornell Box. The global illumination (1) is composed of diffuse direct (2),  diffuse indirect (3), and glossy reflectance (4). }
    \label{fig:cornell-box}
\end{figure}

In \autoref{fig:cornell-box}, we show the lighting decomposition for a single 3D Gaussian model under the Cornell Box lighting, illustrating our pipeline’s classification approach. 

\end{document}